\documentclass[onefignum,onetabnum]{siamart190516} 
\usepackage{lipsum}
\usepackage{amsfonts}
\usepackage{graphicx}
\usepackage{epstopdf}
\usepackage{algorithmic}
\usepackage{mathrsfs}
\usepackage{microtype}
\usepackage{amsfonts}
\usepackage{graphicx}
\usepackage{epstopdf}
\usepackage{amsmath, amssymb}
\usepackage{eufrak} 
\usepackage{yfonts}  
\usepackage[title]{appendix} 
\usepackage{mathtools}   
\usepackage{cleveref}  
\usepackage{enumerate} 
\usepackage{bm}  
\usepackage{soul}  
\usepackage[utf8]{inputenc} 
\usepackage[T1]{fontenc}
\usepackage[normalem]{ulem}
\usepackage{stmaryrd}
\usepackage{tikz}          
\usepackage[]{geometry} 
\numberwithin{equation}{section}
\usepackage{dsfont}
\usepackage{graphicx}   
\usepackage{subcaption}

\ifpdf
  \DeclareGraphicsExtensions{.eps,.pdf,.png,.jpg}
\else
  \DeclareGraphicsExtensions{.eps}
\fi

\usepackage{enumitem}
\setlist[enumerate]{leftmargin=.5in} 
\setlist[itemize]{leftmargin=.5in}

\newsiamremark{remark}{Remark}
\newsiamremark{example}{Example}
\newsiamremark{hypothesis}{Hypothesis}
\crefname{hypothesis}{Hypothesis}{Hypotheses} 
\newsiamthm{claim}{Claim}
\newsiamthm{assumption}{Assumption}

\newcommand{\jt}{\widetilde{\theta}}
\newcommand{\qf}{\mathbf{q}}
\newcommand{\Af}{\mathbf{A}}

\newcommand{\Bf}{\mathbf{B}}
\newcommand{\If}{\mathbf{I}}
\newcommand{\nf}{\mathbf{n}}
\newcommand{\Xo}{\overline{X}}

\newcommand{\zfb}{\overline{\mathbf{z}}}

\newcommand{\oup}{\psi^+}
\newcommand{\uup}{\psi^-}

\newcommand{\ito}{It\^o}

\newcommand{\frU}{\textfrak{U}}
\newcommand{\frp}{\textfrak{p}}

\newcommand{\calM}{\mathcal{M}}
\newcommand{\calL}{\mathcal{L}}
\newcommand{\calK}{\mathcal{K}}
\newcommand{\CST}{\bm{\Delta}}
\newcommand{\clco}{\cl(\mathcal{O})}
\newcommand{\scrA}{\mathscr{A}}
\newcommand{\scrR}{\mathscr{R}}
\newcommand{\scrL}{\mathscr{L}}

\newcommand{\oP}{\overline{P}}
\newcommand{\oM}{\overline{M}}
\newcommand{\oB}{\overline{B}}

\newcommand{\1}{\mathds{1}}

\newcommand{\calF}{\mathcal{F}}

\newcommand{\calB}{\mathcal{B}}
\newcommand{\calA}{\mathcal{A}}
\newcommand{\calS}{\mathcal{S}}
\newcommand{\calJ}{\mathcal{J}}

\newcommand{\frA}{\mathfrak{A}}

\newcommand{\calV}{\mathcal{V}}

\newcommand{\calG}{\mathcal{G}}

\newcommand{\calO}{\mathcal{O}}

\newcommand{\dsN}{\mathbb{N}}  
\newcommand{\dsE}{\mathbb{E}}
\newcommand{\dsF}{\mathbb{F}}

\newcommand{\dsQ}{\mathbb{Q}} 
\newcommand{\dsP}{\mathbb{P}}

\newcommand{\dsR}{\mathbb{R}}

\newcommand{\dsG}{\mathbb{G}}

\newcommand{\SUM}{\displaystyle\sum}

\newcommand{\zf}{\mathbf{z}}

\DeclareMathOperator*{\Tr}{Tr}
\DeclareMathOperator*{\USC}{USC}
\DeclareMathOperator*{\LSC}{LSC}
\DeclareMathOperator*{\cl}{cl}
\usepackage{bbding}
\usepackage{thmtools}

\newcommand*\df{\mathop{}\!\mathrm{d}}

\newcommand{\CJ}[1]{\textcolor[rgb]{1,0., 0.0}{#1}}

\newcommand{\Rom}[1]{\uppercase\expandafter{\romannumeral #1\relax}}
\newcommand{\rom}[1]{\lowercase\expandafter{\romannumeral #1\relax}}

\newcommand{\RN}[1]{%
  \textup{\uppercase\expandafter{\romannumeral#1}}%
}

\DeclareMathOperator{\diag}{diag}

\usepackage{amsopn}

\headers{Lifetime Ruin under HWM Fees and Drift Uncertainty}{J. Lee, X. Yu, and C. Zhou}

\title{Lifetime Ruin under High-watermark Fees and Drift
  Uncertainty\thanks{The first and third authors acknowledge the support from the Singapore MOE AcRF
grants R-146-000-271-112 and R-146-000-255-114 as well as the French Ministry of
Foreign Affairs and the Merlion programme.  The second author is partially
supported by the Hong Kong Early Career Scheme under grant no. 25302116 and the
Hong Kong Polytechnic University central research grant under no.15304317. In
addition, the first author received the financial support from the Singapore MOE
AcRF grant R-146-000-243-114 and the third author received the financial support
from the NSFC Grant 11871364.}  }

\author{Junbeom Lee\thanks{Department of Sales and Trading, Yuanta Securities
    Korea, 04538 Seoul, Korea. Email: \href{junbeoml22@gmail.com}{junbeoml22@gmail.com}}
  \and Xiang Yu\thanks{Department of Applied Mathematics, The Hong Kong
    Polytechnic University, Hung Hom, Kowloon, Hong Kong. Email: \href{xiang.yu@polyu.edu.hk}{xiang.yu@polyu.edu.hk}}
  \and Chao Zhou\thanks{Department of Mathematics, National University of
    Singapore, Singapore 119076, Singapore. Email: \href{matzc@nus.edu.sg}{matzc@nus.edu.sg}}}

\ifpdf
\hypersetup{
  pdftitle={Lifetime Ruin under HWM Fees and Drift Uncertainty}
  pdfauthor={J. Lee, X. Yu, and C. Zhou}
}
\fi

\begin{document}  

\maketitle
    
\begin{abstract}
This paper aims to study lifetime ruin minimization
  problem by considering investment in two hedge funds with high-watermark
  fees and drift uncertainty. Due to multi-dimensional performance fees that are
  charged whenever each fund profit exceeds its historical maximum, the value
  function is expected to be multi-dimensional. New mathematical challenges
  arise as the standard dimension reduction cannot be applied, and the convexity
  of the value function and Isaacs condition may not hold in our probability minimization problem with drift uncertainty. We propose to employ
  the stochastic Perron's method to characterize the value function as the
  unique viscosity solution to the associated \textit{Hamilton–Jacobi–Bellman}
  (HJB) equation without resorting to the proof of dynamic programming
  principle. The required comparison principle is also established in our
  setting to close the loop of stochastic Perron's method.
  

  
\end{abstract}
 
\section{Introduction}
\label{intro}
Hedge funds have existed for many decades in financial markets and have
become increasingly popular in recent times. As opposed to the individual
investment, hedge funds pool capital and invest in a variety of assets and it
is administered by professionals. Hedge fund managers charge performance fees
for their service to individual investors as some regular fees proportional to
fund's component assets plus a fraction of the fund's profits. The most common
scheme entails annual fees of $2\%$ of assets and $20\%$ of fund profit whenever
the profit exceeds its historical maximum---the so-called
\textit{high-watermark}. In the present paper, we are interested in investment
opportunities among several hedge funds and we intend to study a stochastic
control problem given the path-dependent trading frictions as multi-dimensional
high-watermark fees.

The existing research on high-watermark fees mainly has focused on the
asset management problem from the point of view of the fund manager, see some
examples by \cite{Goetzmann}, \cite{Panageas}, \cite{Agar},
\cite{guasoni2016incentives} and \cite{guasoni2015hedge}. Meanwhile, the
high-watermark process is also mathematically related to wealth
drawdown constraints studied in \cite{Grossman}, \cite{CvKa}, \cite{ElieTouzi}
and also discussed in \cite{CYZ} after the transformation into expectation
constraint. Recently, the high-watermark fees have been incorporated 
also into Merton problem for individual investor together with consumption
choice in \cite{janevcek2012optimal} and \cite{JanLiS}. In the presence with
consumption control, analytical solutions can no longer be promised as in some
of the previous work for fund managers. After identifying the state processes,
the path-dependent feature from high-watermark fees can be hidden so
that the dynamic programming argument can be recalled to derive the HJB equation
heuristically. The homogeneity of power utility function in
\cite{janevcek2012optimal} and \cite{JanLiS} enables the key dimension reduction
of the value function and the associated HJB equations can be reduced into ODE
problems. Although the regularity can hardly be expected, classical Perron's
method can be applied and the nice upgrade of regularity of the viscosity
solution can be exercised afterwards using the convexity property of the
transformed one-dimensional value function. As the last step, the verification
theorem can be concluded with the aid of the smoothness of value function and
standard It\^{o} calculus.

In the present paper, we focus on the standpoint of the individual investor who
confronts multiple hedge fund accounts in the market. However, we aim to
minimize the probability that the investor outlives her wealth, also known as
the probability of lifetime ruin, instead of the Merton problem on portfolio or
consumption. We determine the optimal investment strategy of an individual among
some hedge funds who targets a given rate of consumption by minimizing the
probability that the ruin occurs before the death time. For the studies of
lifetime ruin probability problem, readers can refer to
\cite{young2004optimal,bayraktar2007minimizing,bayraktar2015stochastic,bayraktar2015minimizing,young2016lifetime}. In
contrast to Merton problem, the dimension reduction of  
the value function will fail for our probability minimization problem. The  
auxiliary controlled state process, the so-called process of distance to pay
performance fees defined in \eqref{def.Y}, can no longer be absorbed to simplify
the PDE problem. Furthermore, comparing with \cite{janevcek2012optimal} and
\cite{JanLiS} or the lifetime ruin problem with ambiguity aversion in
\cite{bayraktar2015minimizing}, we need to handle a genuine multi-dimensional
control problem with reflections as there exist multiple hedge funds in the
market. In other words, the distance process itself is already
multi-dimensional, which spurs many new mathematical challenges. To wit, one can
still exploit the classical Perron's method as in \cite{janevcek2012optimal},
\cite{JanLiS} and \cite{bayraktar2015minimizing}, and obtain the existence of
viscosity solution to the associated HJB equation. Nevertheless, the upgrade of
regularity of the viscosity solution can hardly be attained for our
multi-dimensional problem. Consequently, the proof of verification theorem,
which requires certain regularity of the solution, cannot be completed. To
relate the value function to the viscosity solution in our setting using
classical Perron's method, we have to provide the technical proof of dynamic
programming principle at the beginning.

In addition, the individual investor usually cannot keep a real-time track of
the performance of hedge funds from fund managers. Moreover, a reliable
estimation of the return from hedge fund that consists of a bunch of various
assets is almost impossible in practice. Even in the hedge fund performance
report, the predicted future return in short term from fund manager is provided
as a certain range instead of a fixed number. It is more realistic to assume
that the investor allows drift misspecification and starts with a family of
plausible probability measures of the underlying model. This leads to a robust
investment strategy with Knightian model uncertainty. In particular, we assume
that the investor would like to use the available data as a reference model and
work on a robust control problem with the penalty on other plausible models
based on the deviation from the reference one. One new mathematical challenge
from this formulation is that the value function may lose convexity for some
parameters and the Issacs condition may fail. Adding our previous difficulties
coming from multi-dimensional performance fees, the feedback optimal investment
strategy and the saddle point choice of probability measure cannot be
obtained. The combination of market imperfections such as trading frictions
together with model ambiguity renders many problems mathematically
intractable. Some workable examples in this direction can only be found in
robust Merton problem with proportional transaction costs, see \cite{NeuSik},
\cite{ChauRas} and \cite{dengtanyu}. The methodology introduced in these paper
may not work for our purpose with path-dependent high-watermark fees.

To tackle our stochastic control problem, we choose to employ the stochastic
Perron's method (SPM) and characterize the value function as the unique
viscosity solution to the associated HJB equation. This stochastic version of
Perron's method, introduced by \cite{bayraktar2012stochastic}, can avoid the
technical and lengthy proof of dynamic programming principle (DPP) and can
obtain it as a by-product. We choose SPM over the weak DPP introduced in
\cite{bouchard2011weak} because SPM can better handle the path-dependent
structure of our control problem with additional model uncertainty. Let us note
that the comparison principle is needed anyway in both methods. SPM requires the
comparison principle to complete the squeeze argument and establish the
equivalence between value function and the viscosity solution, while weak DPP
needs the comparison principle to guarantee the uniqueness of the viscosity
solution to the associated HJB equation. We actually find that the proof of
comparison principle for SPM is relatively easier as the applicable class of
state processes can be larger than that of weak DPP.
We refer a short list of previous work on stochastic control using SPM such as \cite{bayraktar2012stochastic}, \cite{bayraktar2014stochastic}, \cite{bayraktar2016stochastic}, \cite{bayraktar2017controller}, \cite{bayraktar2013stochastic}, \cite{rokhlin2014verification}, \cite{Sirb1}, \cite{Sirb2}, \cite{BayCossPham} and \cite{YangYu}.

To establish the viscosity semisolution property of stochastic envelopes, it is usually
crucial to check the boundary viscosity semisolution property. In our framework,
we can take advantage of the problem structure from lifetime ruin probability minimization and explicitly construct a
\textit{stochastic super-solution} and a \textit{stochastic sub-solution} which satisfy the desired
boundary conditions. We note that our arguments using stochastic Perron's method
differ from \cite{bayraktar2015stochastic} that solves the lifetime ruin problem
with transaction costs and \cite{BayCossPham} that examines the robust optimal
switching problem. Some nontrivial issues need to be carefully addressed, which
are caused by the uncertainty of drift term and the structure of the auxiliary
state process defined as the distance to pay fees. The path-dependent running
maximum part coming from high-watermark fees do not appear in
\cite{bayraktar2015stochastic} nor \cite{BayCossPham}, which deserves some novel
and tailor-made treatment in the present paper.

It is the scope of this paper to investigate a multi-dimensional stochastic
control problem on the strength of stochastic Perron's method, which integrates
the drift ambiguity and high-watermark fees from multiple hedge funds. The
generality of the mathematical problem comes at the cost that the associated HJB
equation becomes numerically challenging. First, our HJB equation naturally has
three spatial variables and a dimension reduction technique
cannot be applied to our objective
function.
In addition, due to the nature of ruin probability mimization and the
high-water mark fees, both Dirichlet and Neumann boundary conditions are imposed
for our HJB equation. It is well known that the stability and efficiency of
numerical schemes may become big issues for the high dimensional nonlinear PDE
with mixed type boundary conditions. The numerical analysis and the study of
quantitative impacts by high-water mark fees and parameter uncertainty will be
pursued in our future research. It will be interesting to apply the deep
learning method in the future work to tackle our multi-dimensional nonlinear PDE
with mixed boundary conditions as in \cite{sirignano2018dgm}.

The rest of the paper is organized as follows. Section \ref{sec-2} introduces
the market model with multiple hedge funds and related high-watermark fees, the
default time as well as the set up with drift uncertainty. The robust lifetime
ruin problem is defined afterwards. In Section \ref{sec-3}, we derive the
associated HJB equation for the control problem heuristically and define the viscosity
solution accordingly. The main theorem to characterize the value function as the
unique viscosity solution is presented. Section \ref{sec-4} provides the proof
of all main results using stochastic Perron's method. The proof of the
comparison principle of the HJB equation is also reported therein.
 
\section{Market Model and Problem Formulation}\label{sec-2}
\subsection{Multiple Hedge Funds with High-watermark Fees}
Let $(\Omega, \calG, \dsG, \dsP)$ be a filtered probability space such that
$\dsG$ satisfies the \textit{usual conditions} and $\dsE$ denote the expectation
operator under $\dsP$. Let $(W_t)_{t\geq0}$ denote an independent
$2$-dimensional Brownian motion and $\dsF\coloneqq (\calF_t)_{t \geq0}$ be the
\textit{natural filtration} generated by $(W_t)_{t\geq0}$ and it is assumed that
$\calF_t\subset\calG_t$. Later, we will characterize $\dsG=(\calG_t)_{t\geq 0}$ more
precisely.

We consider the financial market consisting of one risk-less bond with interest
rate $r> 0$ and two hedge fund accounts $(F^i_t)_{t\geq0}$, $i\in\{1, 2\}$, described
by
\begin{align}
  \df F^i_t=& \mu^i F^i_t\df t + \sigma ^iF^i_t \df W_t,\nonumber
\end{align}
for some constant $\mu^i\geq0$ and constant vector $\sigma^i \in \dsR^{2}$. To simplify the presentation, we only focus on two hedge funds henceforth. The mathematical arguments and main results can be easily extended to the multi-dimensional case of $N\geq 2$ hedge funds without
any technical difficulty. 
We shall denote 
\begin{align}
  F\coloneqq
  \begin{bmatrix}
 F^1\\ F^2   
\end{bmatrix}, \quad
  \mu \coloneqq
  \begin{bmatrix}
    \mu^1\\\mu^2
  \end{bmatrix}, \quad
  \sigma \coloneqq
  \begin{bmatrix}
    \sigma^1\\
    \sigma^2
  \end{bmatrix},\nonumber
\end{align}
and assume that $\sigma$ is invertible. 

Contrary to some standard investment problems in liquid risky assets such as
stocks, we are considering the model when the investor is facing the wealth
allocation among some hedge fund accounts that charge proportional fees on the
profit as trading frictions. In particular, the investor needs to pay some
high-watermark  fees to the fund manager whenever the accumulative
profit reaches the highest value. The 2/20-rule is common for hedge funds in the
sense that $2\%$ per year of the total investment and $20\%$ of the additional
profits are paid to the fund manager whenever the high-watermark exceeds the
previously attained profit maximum. To explain this in a more explicit manner,
let $\pi=(\pi^1, \pi^2)\in \dsR^{2}$ denote the investment strategy in two hedge funds
$F$.  The accumulative profit $\oP^\pi = [\oP^{1, \pi}, \oP^{2, \pi}]^\top$ from the
hedge fund before the deduction of the high-watermark fee, is characterized by
the stochastic integral
\begin{align}
  \oP^{i, \pi}_t  \coloneqq& \int_{0}^{t}\pi^i_s \frac{\df
                           F^i_s}{F^i_s}. \label{def.pb} 
\end{align}
In practice, the investor and fund manager may agree to choose a
  \textit{benchmark} to measure the manager's performance, see
  \cite{JanLiS}. High-watermark fees are only deducted when the profit process of
  the fund exceeds the benchmark level. For example, the fund manager may only receive incentives when the fund account outperforms the S\&P index.

The initial high-watermark fee is denoted by some non-negative
constant vector $y = [y^1, y^2]^\top$. Let $F^B\in \dsR^2$ be the \textit{benchmark}
process given by
\begin{align}
 \df F^B_t =\diag(F^B_t)[\mu^B \df t + \sigma^B \df W_t],\nonumber
\end{align}
for some $\mu^B\in \dsR^2$, $\sigma^B \in \dsR^{2\times 2}$. We denote by
$\oB^\pi = [\oB^{1, \pi}, \oB^{2, \pi}]^\top$ the accumulated \textit{benchmark} profit
process if the same strategy $\pi$ is adopted, i.e.,
\begin{align}
 \oB^{i, \pi}_t \coloneqq \int_{0}^{t} \pi^i_s \frac{\df F^B_s}{F^B_s}.\nonumber
\end{align}
Let $q = [q^1, q^2]^\top$ represent the proportional rates of high-watermark fee of
each hedge fund and $P^{y, \pi}=[P^{1, y, \pi}, P^{2, y, \pi}]$ be the realized profit
after charging the high-watermark fee. Moreover, we define $M^{i,y,\pi}$ as the
historical high-watermark of the $i$-th hedge fund. The realized profit process
$P^{i, y, \pi}$, $i\in \{1, 2\}$, is given by
\begin{align}
\begin{cases}
\df P^{i, y, \pi}_t \coloneqq  \df \oP_t^{i, \pi} - q^i \df M^{i, y, \pi}_t,
  &P^{i, y, \pi}_0 =0,\label{psde.crude}\\
  M_t^{i, y, \pi} \coloneqq  \sup_{0\leq s\leq t}\big\{(P^{i, y, \pi}_s - \oB^{i, \pi}_s) \vee y^i\big\},
  ~~&y^i\geq0.    
\end{cases}
\end{align}
To represent
 \cref{psde.crude}  
in a more convenient form, let us define
\begin{align}
 \overline{M}_t^{i, y, \pi} \coloneqq& \sup_{0\leq s\leq t}\big\{(\oP^{i, \pi}_s - \oB^{i,
                                     \pi}_s) \vee y^i\big\}, ~~i \in \{1, 2\}.
                                  \label{def.mb}
\end{align} 
Then by \cref{psde.crude}, 
\begin{align}
 \oM_t^{i, y, \pi} -y^i=& \sup_{0 \leq s \leq t}\big\{[\oP^{i, \pi}_s - \oB^{i, \pi}_s]-y^i\big\}^+\nonumber\\
=&\sup_{0 \leq s< t} \big\{[P^{i, y, \pi}_s - \oB^{i, \pi}_s]-y^i + q^i[M^{i, y, \pi}_s-y^i]\big\}^+\nonumber\\
=&(1+q^i)(M^{i, y, \pi}_t-y^i). \label{sde.Mb}
\end{align}
Therefore, in view of \cref{psde.crude} and \cref{sde.Mb}, for $i \in \{1, 2\}$,
we have
\begin{align}
  P^{i, y, \pi}_t =& \oP_t^{i, \pi} - \frac{q^i}{1+q^i} [\oM^{i, y, \pi}_t -y^i],
                           \label{def.p}
\end{align}
Equivalently, $P^{i,y,\pi}$ can be rewritten as
\begin{align}
  \df P^{i, y, \pi}_t  =  \mu^i \pi^i_t\df t + \sigma^i\pi^i_t \df W_t - q^i(1+q^i)^{-1}\df
  \oM_t^{i, y, \pi}. 
\end{align}
As the high-watermark fee is only deducted whenever $\oM^i  -[\oP^i - \oB^i]
=0$, the distance between $\oM^i$ and $\oP^i - \oB^i$ will be considered in the
investment decision. Therefore, let us introduce the distance process $Y^{y, \pi} = [Y^{1, y, \pi}, Y^{2, y, \pi}]^\top$ as the difference
\begin{align}
  Y^{i, y, \pi} \coloneqq
  &M^{i, y, \pi} - [P^{i, y, \pi} - \oB^{i, \pi}]. \label{def.Y}
\end{align}
In view of \cref{sde.Mb}, \cref{def.p}, and \cref{def.Y}, it clearly follows
that $Y^{y, \pi} = \oM^{y, \pi} - [\oP^\pi - \oB^\pi]$. To facilitate the future
analysis using dynamic programming argument, we expect to deal with a multi-dimensional value function of the control problem depending on the two
dimensional initial distance $Y_0=(y^1,y^2)$ and the investor's initial wealth
$x$. The precise formulation will be introduced later.

We continue to characterize the investor's wealth more explicitly. The amount of the risky position (hedge funds) is $\1^\top\pi$ and the rest of the
investor's wealth is put into the risk-less bond. Furthermore, it is assumed that the investor consumes at a constant rate
$c \geq0$ all the time. Let $X^{x, y, \pi}$ denote the process of investor's wealth with initial
value $x$. Then the controlled state processes are given by
\begin{align}\label{def.XY}
\begin{cases}
  \df X^{x, y, \pi}_t = [rX^{x, y, \pi}_t-c+\pi^\top_t\mu^r_\Delta ] \df t +  \pi^\top_t \sigma\df W_t
  -  q^\top\df M^{y, \pi}_t, ~~&X_0 = x,\\
  \df Y^{y, \pi}_t =-\diag(\pi_t)[\mu^{B}_\Delta  \df t  +\sigma^{B}_\Delta  \df W_t] + \diag(\1+q)\df
             M^{y, \pi}_t,~~&Y_0=y.
\end{cases}
\end{align}
where we denote $\mu^r_\Delta\coloneqq [\mu^1-r, ~\mu^2-r]^\top$, $\mu_\Delta^B\coloneqq \mu-\mu^B$,
$\sigma_\Delta^B\coloneqq \sigma-\sigma^B$, and 
\begin{align}
\diag(\1+q)\coloneqq
  \begin{bmatrix}
    1+q^1 & 0\\
    0 & 1+q^2
  \end{bmatrix}.\nonumber
\end{align}
Sometimes, we omit the superscripts $x, y, \pi$ for simplicity and  we also denote
\begin{align}
  Z\coloneqq (X, Y^1, Y^2), \quad \zf \coloneqq (x, y^1, y^2).\nonumber
\end{align}

\subsection{Default Time and Preliminaries}
Another important ingredient of our model is the default time of the individual
investor, such as the death time independent with $(W_t)_{t\geq0}$, which is defined as a random variable 
\begin{align}
\tau_D: (\Omega, \calG) \rightarrow (\dsR_+, \textfrak{B}(\dsR_+))    \nonumber 
\end{align}
satisfying $\dsP(\tau_D = 0) = 0$ and $\dsP(\tau_D > t) >0$, 
for any $t \geq0$.  From this point onward, the \textit{full market filtration} $\dsG$ is precisely defined by
 $\dsG =(\calG_t)_{t \geq0}
 \coloneqq  (\calF_t \vee \sigma(\{\tau_D \leq u\}: u \leq t))_{t \geq0} $.
It is worth noting that $\tau_D$ is a $\dsG$-stopping time but may fail to be an $\dsF$-stopping
time. In what follows, we assume that there exists a constant $\lambda^D  >0$ such that
\begin{align}
G^D_t\coloneqq \dsP(\tau_D >t \vert \calF_t) = e^{-\lambda^D t}. \nonumber
\end{align}
We call $\lambda^D$ the intensity of default time $\tau_D$ with respect to
$\dsF$. Under this assumption,
\begin{align}
 (\calM_t^D)_{t\geq0} \coloneqq \big(\1_{\tau_D \leq t}- \lambda^D(t \wedge
  \tau_D)\big)_{t\geq 0}
\end{align}
is a  $(\dsG)$-martingale. Moreover, for any $\dsF$-martingale
$(\xi_t)_{t\geq0}$, $(\xi_{t\wedge \tau^D})_{t\geq0}$ is a
$\dsG$-martingale. Therefore, $(W_t)_{t\geq 0}$ is a $\dsG$-Brownian motion; see
\cite{elliott2000models}.      
\begin{remark}
\begin{enumerate}[label=\arabic*.]
\item In view of the existence of the intensity, $\tau_D$ is \textit{totally
    inaccessible}. In other words, the default of the investor comes with total
  surprise. On the other hand, a \textit{ruin time}, which will be introduced
  later, is defined as a hitting time that the controlled wealth process crosses
  a given level and it is therefore \textit{predictable}. In the present paper,
  we envision an individual investor who chooses her portfolio to minimize the
  probability involving the \textit{ruin time} before the default time occurs.
  
\item Although investment strategies are defined as $\dsG$-adapted processes, the
  \textit{full filtration} $\dsG$ is not fully observable for the
  investor. However, in this filtration setup, for any $\dsG$-adapted process,
  we can find an $\dsF$-reduction, where $\dsF$ is the observable
  information. Therefore, the strictly $\dsG$-adapted strategies only describe
  an immediate action taken by the investor at the default time. Note that an
  $\dsF$-adapted process is not necessarily determined independently of the
  default time $\tau_D$, because the (constant) default intensity $\lambda^D$ is
  trivially $\dsF$-adapted.
\end{enumerate}
\end{remark}

\subsection{Life Time Ruin Problem with Drift Uncertainty}
Based on previous building blocks, we are ready to introduce the primary
stochastic control problem that the investor confronts. In particular, the
investor concerns the viability of her investment before the default time and she
wishes to maintain the amount of her wealth above a certain level, say
$R \geq0$, before the default time happens. To this end, it is natural to introduce
the so-called ruin time
\begin{align}
  \tau_R^{x, y, \pi}\coloneqq \inf\{t \geq0~\colon X^{x, y, \pi}_t \leq R~\}.\nonumber
\end{align}
Mathematically speaking, the investor chooses $\pi$ from an admissible set $\scrA$
so that $\tau_R$ occurs as 
late as possible. As the investor cannot control the \textit{totally
  inaccessible time} $\tau_D$, she aims to minimize the probability that the ruin occurs before
the default time.

However, we consider a more practical scenario in the present paper that the
return of hedge funds may not be revealed by fund manager to the investor very
frequently. The investor usually can only get access to the performance of the
fund from some reports on regular dates. Moreover, as the hedge fund consists of
components from various assets, the estimation of return can hardly be provided
on a timely basis. Based on these observations, it is reasonable to assume that
the investor may not have a precise knowledge of the dynamics of hedge
funds. This naturally leads to the so-called Knightian model uncertainty.

In this paper, we will only focus on the case with drift uncertainty, i.e. the
investor conceives a family of plausible return terms from the hedge fund
dynamics and proceeds to solve the control problem in a robust sense. Indeed,
the precise estimation of the drift term is much more challenging than the
estimation of volatility term, which motivates our research. In particular, we
aim to minimize the probability of lifetime ruin by choosing wealth allocation
among multiple hedge funds with high-watermark fees and drift uncertainty, which
is new to the existing literature. To this end, let us first introduce a class
of probability measures equivalent to the reference probability $\dsP$
and denote this class by $\scrL$.
\begin{definition}\label{def.theta}
$\dsQ\in \mathscr{L}$ if for any $0\leq t$,
\begin{align}
  \frac{\df\dsQ}{\df \dsP}\bigg|_{\calG_t}
  =\exp\bigg(-\frac{1}{2}\int_{0}^{t}\Vert \theta_s\Vert ^2 \df s
  +\int_{0}^{t}\theta_s^\top\df W_s\bigg), \label{def.Q}
\end{align}
for some $\dsG$-predictable process $\theta$ valued in a closed set
$\calL \subseteq \dsR^2$ containing $\bf{0}$ such that
\begin{eqnarray}
  &\dsE^{\dsQ}\Big[\int_{0}^{\infty}e^{-\lambda^D s}
    \Vert \theta_s\Vert^2 \df s\Big] <\infty, \nonumber\\
  &\dsE\Big[\exp{\Big(\frac{1}{2} \int_{0}^{t}
    \Vert \theta_s\Vert^2 \df s\Big)}\Big] <\infty,~~\text{for any } t \geq
    0. \nonumber
\end{eqnarray}
\end{definition}
In what follows, an equivalent measure $\dsQ$ is generated by $\theta$ by the
representation in \cref{def.Q}, and we call $\dsQ$ the $\theta$-measure. The investor
intends to minimize the ruin probability under some $\dsQ \in \scrL$, but the
deviation of the measure from $\dsP$ is penalized by a relative entropy process
up to the default time $\tau_D$:
\begin{align}\label{entp}
    H_t(\dsQ|\dsP)\coloneqq
  \dsE^\dsQ\Big[\log\Big(\frac{\df\dsQ}{\df\dsP}\Big|_{\calG_t}\Big)\Big],
  ~~\text{for } t \geq0.
\end{align}

The investor's robust stochastic control problem is then defined by
\begin{align}
  V(x, y; \varepsilon)\coloneqq \inf_{\pi \in \mathscr{A}}\sup_{\dsQ\in \mathscr{L}}
  \Big\{\dsQ(\tau^{x, y, \pi}_R < \tau_D) - \frac{1}{\varepsilon}H_{\tau_D}(\dsQ\mid\dsP)\Big\}.\label{def.value}
\end{align}
Here $\scrA$ denotes the set of all admissible controls defined in the following sense.
\begin{definition}
 $\pi\in \scrA$ if $\pi$ is $\dsG$-predictable and valued in a compact set $\calK\subseteq
 \dsR^{2}$ such that $(0, 0)\in \calK$. 
\end{definition}

\begin{remark} The coefficient $\varepsilon$ in the penalty term of \eqref{def.value} corresponds to the investor's level of model ambiguity about the reference probability $\mathbb{P}$. For instance, the case $\varepsilon\rightarrow 0$ implies that 
\begin{align*}
\sup_{\dsQ\in \mathscr{L}}
  \Big\{\dsQ(\tau^{x, y, \pi}_R < \tau_D) - \frac{1}{\varepsilon}H_{\tau_D}(\dsQ\mid\dsP)\Big\}\rightarrow
  \dsP(\tau^{x, y, \pi}_R < \tau_D),\end{align*}
which indicates that the investor is completely confident about the probability measure $\mathbb{P}$. On the other hand, if the agent is extremely uncertain as $\varepsilon\rightarrow\infty$, we get that
\begin{align*}
  \sup_{\dsQ\in \mathscr{L}}
  \Big\{\dsQ(\tau^{x, y, \pi}_R < \tau_D) - \frac{1}{\varepsilon}H_{\tau_D}(\dsQ\mid\dsP)\Big\}\rightarrow
  \sup_{\dsQ\in \mathscr{L}}
  \dsQ(\tau^{x, y, \pi}_R < \tau_D),\end{align*}
which reduces to the worst-case scenario. It is worth noting that the
formulation involving the penalty term only works for drift uncertainty. If
some plausible probabilities are mutually  singular due to volatility
uncertainty, i.e. there is no dominating reference probability $\mathbb{P}$,
the entropy cannot be defined as in  \eqref{entp}. Another interesting issue we
can consider in the robust framework is to incorporate the investor's ambiguity
attitude towards a given set of plausible priors. Similar to \cite{huangyu},
one can employ the alpha-maxmin preference and formulate the ruin probability problem under model uncertainty as
\begin{align*}
  \inf_{\pi \in \mathscr{A}} \left[\alpha\sup_{\dsQ\in \mathscr{L}}
  \dsQ(\tau^{x, y, \pi}_R < \tau_D) + (1-\alpha)\inf_{\dsQ\in \mathscr{L}}
  \dsQ(\tau^{x, y, \pi}_R < \tau_D)\right].\end{align*} 
This formulation allows for both drift and volatility uncertainty and the
constant coefficient
$\alpha\in[0,1]$ can represent how much ambiguity averse the investor is. Nevertheless, this problem becomes time inconsistent and we need to look for some equilibrium portfolio strategies instead of the optimal one, which is beyond the scope of this paper and will be left as future research.
\end{remark} 
 
\begin{remark} 
  The compactness of $\calK$ in the definition of admissible set $\scrA$ can be understood that the investor does not take
  an extreme strategy and the immediate liquidation is also
  admissible. Moreover, as $\pi$ is $\dsG$-predictable, it is also
  $\dsF$-predictable before $\tau_D$. Therefore, there is a unique continuous
  $\oP^\pi$ satisfying \cref{def.pb}. Thanks to \cref{def.mb} and \cref{def.p},
  $P^{y, \pi}$ is well-defined.  More importantly, the compactness of $\calK$ is
  necessary for the associated HJB equation to be
  continuous. Otherwise, it becomes difficult to prove the comparison principle for its
  viscosity solutions because the typical doubling argument relies on Crandall-Ishii's
  lemma and the closure of \textit{super/sub-jets}, which require the compactness of $\calK$. In other
  words, if the comparison principle is already guaranteed, we can relax the
  conditions on $\scrA$ only with care for $P^{y, \pi}$ to be well-defined. 
\end{remark}
\begin{remark}
  \begin{enumerate}
  \item One can naturally generalize our model to include ambiguity on the
    hazard rate as well. Nevertheless, the additional ambiguity on default time
    does not complicate our analysis on the associated HJB equation and our
    methodology still holds valid. For a related work on life time ruin problem
    with uncertain hazard rate (but without high-watermark fees), we refer to
    \cite{young2016lifetime}, in which the one-dimensional HJB equation can be
    solved by a verification argument.
    \item The main mathematical challenge comes from the
        multi-dimensional high-water mark fees. Even without drift uncertainty,
        our stochastic control problem is still three dimensional together with
        mixed boundary conditions, which does not admit any closed form
        solution. The examination of the impact by the uncertainty parameter
        $\varepsilon$ in our model would be appealing, which nevertheless relies
        on some stable and efficient numerical schemes. As some conventional
        numerical methods may not work well for our multi-dimensional nonlinear
        PDE with mixed boundary conditions, we will not explore this direction
        further in the left of the paper and leave the numerical treatment and
        sensitivity analysis as future work.
  
    
  \end{enumerate}

\end{remark}
\section{Dynamic Programming Equation and Main Results}\label{sec-3}
In this section, we first heuristically derive the HJB equation associated with the value
function using dynamic programming argument or martingale optimality principle. For technical reason, when default occurs, we assign a coffin state $\CST$
to the underlying process $Z$. Moreover,  for any domain in what follows, we
consider its one point compactification and any function $u$ is extended by
assigning $u(\CST)=0$. 
Denote the $(\dsQ, \dsG)$-Brownian motion by $W^\dsQ$, where $\dsQ$ is
generated by $\theta$. For $t < \tau_D$, \cref{def.XY} can be written as
\begin{align}
\begin{cases}
 \df X^{x, y, \pi}_t = [rX^{x, y, \pi}_t-c+\pi^\top_t(\mu^r_\Delta+\sigma\theta) ] \df t +  \pi^\top_t \sigma\df
      W^\dsQ_t -  q^\top\df M^{y, \pi}_t, ~~&X_0 = x, \\
  \df Y^{y, \pi}_t =-\diag(\pi_t)[(\mu^{B}_\Delta +\sigma^B_\Delta\theta) \df t  +\sigma^{B}_\Delta  \df W^\dsQ_t] 
+ \diag(\1+q)\df M^{y, \pi}_t,~~&Y_0=y. 
\end{cases}
\end{align}
To obtain the associated HJB equation, we apply \ito's formula to a smooth
function $\varphi$ that
\begin{align}
  \df \varphi(Z_t) -\frac{1}{2\varepsilon}\Vert\theta_t\Vert^2\df t=
  &\big[- \lambda^D[\varphi(Z_t) - \varphi(\CST)] +
    (rX_t-c)\varphi_x +\calA^{\pi_t, \theta_t}[\varphi] (Z_{t})\big]\df t
    \nonumber\\
  &-\SUM_{i=1,2}\big[q^i
    \varphi_x(Z_t)-(1+q^i)\varphi_{y^i}(Z_t)\big]\1_{Y^i_t=0}\df M^i_t
    \nonumber\\
  & +\big[\varphi_x(Z_{t})\pi^\top_t\sigma - \nabla_y\varphi(Z_{t})^\top\diag(\pi_t)\sigma^B_\Delta\big] \df W^\dsQ_t
    +[\varphi(\CST)-\varphi(Z_{t-})]\df \calM^D_t,
    \label{ito.dpe}
\end{align} 
where $\nabla_y\varphi \coloneqq [\partial_{y^1}\varphi,~ \partial_{y^2}\varphi]^\top$ and
\begin{align}
  \calA^{\pi, \theta} [\varphi](x, y^1, y^2) \coloneqq &  -\frac{1}{2\varepsilon}\Vert\theta\Vert^2+b[\pi, \theta]^\top\nabla \varphi + \frac{1}{2}\Tr(\Sigma[\pi]\nabla^2\varphi),\nonumber\\
  b[\pi, \theta]\coloneqq
  & \begin{bmatrix}
       \pi^\top(\mu^r_\Delta + \sigma\theta) \\
       -\diag(\pi)(\mu^{B}_\Delta +\sigma^B_\Delta\theta) 
     \end{bmatrix},\nonumber\\
  \Sigma[\pi]\coloneqq
  &
    \begin{bmatrix}
      \pi^\top\sigma\\
      -\diag(\pi)\sigma^B_\Delta
    \end{bmatrix}\begin{bmatrix}
      \pi^\top\sigma\\
      -\diag(\pi)\sigma^B_\Delta
    \end{bmatrix}^\top.\nonumber
\end{align}
Recall that $\varphi(\CST)=0$ in \cref{ito.dpe}. Now, let us deduce related boundary
conditions. Recalling \cref{def.value}, we can set  $V(R, y^1, y^2) =1$
for any $y^i \geq0$. In addition, if $X_t = c/r$ at $t \geq0$, the optimal strategy is
liquidating the risky position so that $X_s = c/r$ for any $s \geq t$. Therefore,
$V(c/r, y^1, y^2)=0$ for any $y^i \geq0$. Thus, motivated by these boundary
conditions, we need to consider the following regions and boundaries 
\begin{align}
  \calO  \coloneqq& \{(x, y^1, y^2)\colon R<x<c/r,~ y^1\geq0, ~y^2 \geq0\},\nonumber\\
  \calO^+ \coloneqq& \{(x, y^1, y^2)\colon R<x<c/r,~ y^1 >0, ~y^2>0\}, \nonumber\\
  \partial\calO^0_i \coloneqq& \{(x, y^1, y^2)\in \calO\colon R<x<c/r,~ y^i =0\},\nonumber\\
  \partial\calO^0 \coloneqq& \{(x, y^1, y^2)\in \calO\colon R<x<c/r,~ y^1=0 \text{ or }y^2 =0\},\nonumber\\
  \partial\calO_R\coloneqq &  \{(R, y^1, y^2)\colon ~ y^1 >0, ~y^2>0\},\nonumber\\
  \partial\calO_{c/r}\coloneqq & \{(c/r, y^1, y^2)\colon ~ y^1 >0, ~y^2>0\}.\nonumber
\end{align}
Note that $\calO = \calO^+ \cup \partial\calO^0$,
$\partial\calO=\partial\calO_R\cup\partial\calO_{c/r}\cup \partial\calO^0_1 \cup \partial\calO^0_2$, and
$\partial\calO^0 = \partial\calO^0_1 \cup\partial\calO^0_2$. Moreover, for any set
$A$, we let $\cl{(A)}$ denote the closure of $A$ in what follows. We then
consider the following operators
\begin{align}\label{operator.1st}
  \begin{cases}
  \calF[\varphi](\zf)
  \coloneqq\lambda^D\varphi(\zf) -(rx-c)\varphi_x(\zf)-\inf\limits_{\pi\in \calK}
             \sup\limits_{\theta\in \calL}\calA^{\pi, \theta}[\varphi](\zf), \\
 \calB^i[\varphi](\zf)\coloneqq q^i\varphi_x(\zf) - (1+q^i)\varphi_{y^i}(\zf),
                                 ~~i \in \{1, 2\},
  \end{cases}
\end{align}
and the associated HJB
equation can be (formally) written as
\begin{align}\label{dpe.crude}
\begin{cases}
\calF[\varphi](\zf)=0 ,~~&\text{on}~~\zf \in \calO^+, \\
\calB^1[\varphi](\zf) =0, ~~&\text{on}~~\zf\in \partial\calO^0_1, \\
\calB^2[\varphi](\zf) =0, ~~& \text{on} ~~\zf\in \partial\calO^0_2, \\
\varphi(\zf)=1,~~&\text{on} ~~\zf\in \partial\calO_R,\\
 \varphi(\zf)=0,~~&\text{on} ~~\zf\in \partial\calO_{c/r}.    
\end{cases}
\end{align}
\begin{remark}
One may want to solve a benchmark case without uncertainty, namely $\calL
=\{\bf{0}\}$. In this case, while $\calF[\varphi]$ becomes simpler as
\begin{align}
  \calF^0[\varphi](\zf)
  \coloneqq\lambda^D\varphi(\zf) -(rx-c)\varphi_x(\zf)-\inf\limits_{\pi\in \calK}
             \calA^{\pi, \bf{0}}[\varphi](\zf),
\end{align}
the boundary condition $\calB^i$, $i=1, 2$, still remain unchanged. Note that
the major difficulties of our problem are the high dimensionality and the
Neumann-type boundary conditions. Thus, considering the benchmark case does not
provide an easier problem, and classical solution still cannot be
proved. Instead, we will solve the general problem \cref{dpe.crude} using the
stochastic Perron's method in the next section. Note that our mathematical
arguments based on stochastic perron's method for the model with drift
uncertainty can be easily modified to cover the simpler benchmark case without
model uncertainty. It is our goal to provide a streamlined proof for the general
model in the present paper, which is motivated by some practical ambiguous
returns in hedge fund investment.
\end{remark}
Our ultimate goal is to show that the value function $V$ defined in
\eqref{def.value} is the unique viscosity solution of the HJB equation
\cref{dpe.crude}.
To this end, we first need to be careful for the boundary conditions on
$\partial\calO^0$, which should be defined using semi-continuous envelope of
viscosity solutions. To be precise, we denote the lower (resp. upper)
semi-continuous envelope of $\calB^i$, $i \in \{1, 2\}$, by $\calB_*$
(resp. $\calB^*)$. On $\partial\calO^0$, we will consider
\begin{align}
  \calB_*[\varphi]\coloneqq
  \begin{cases}
    \calB^1[\varphi], &\text{ on } \partial\calO^0_1 \setminus \partial\calO^0_2,\\
    \calB^2[\varphi], &\text{ on } \partial\calO^0_2 \setminus \partial\calO^0_1,\\
    \min\{\calB^1[\varphi], ~\calB^2[\varphi]\}, ~&\text{ on } \partial\calO^0_1 \cap \partial\calO^0_2,
  \end{cases}\nonumber
\end{align}
and $\calB^*$ is defined in the same way by replacing $\calB_* = \min\{\calB^1,
\calB^2\}$ using $\calB^* = \max\{\calB^1, \calB^2\}$ on the boundary
 $\partial\calO^0_1 \cap \partial\calO^0_2$. Furthermore, we denote 
\begin{align}
  \text{USC}_b(A) \coloneqq& \{\text{bounded u.s.c functions on } A \},\nonumber\\
  \text{LSC}_b(A) \coloneqq& \{\text{bounded l.s.c functions on } A \}. \nonumber
\end{align}
The precise definition of viscosity sub/super solutions is given as below.
\begin{definition}[Viscosity solution]\label{def.viscosity}
\begin{enumerate}[label=(\roman*)] 
\item $v \in \USC_b(\clco)$ is a  viscosity sub-solution of
(\ref{dpe.crude}) if for any test function $\varphi$ such that $ \zf \in \calO$ is
a strict
maximum point of $v - \varphi$ at zero, we have
\begin{align}
\begin{cases}
 \calF[\varphi](\zf)\leq0, ~&\text{ on }~
     \zf \in \calO^+, \\
\min\big\{\calF[\varphi](\zf), ~\calB_*[\varphi](\zf)\big\} \leq0,
 ~&\text{ on } ~\zf\in \partial\calO^0,
      \\
      v(\zf)\leq1,~&\text{ on } ~\zf\in \partial\calO_R,\\
   v(\zf)\leq0, ~&\text{ on } ~\zf\in \partial\calO_{c/r}.   
\end{cases} 
\end{align}
\item  $v \in \LSC_b(\clco)$ is a  viscosity super-solution of
(\ref{dpe.crude}) if for any test function $\varphi$ such that $ \zf \in \calO$ is a
strict minimum point of $v - \varphi$ at zero, we have
\begin{align}
\begin{cases}
 \calF[\varphi](\zf)\geq0, ~&\text{ on }~
     \zf \in \calO^+, \\
\max\big\{\calF[\varphi](\zf), ~\calB^*[\varphi](\zf)\big\} \geq0,
 ~&\text{ on } ~\zf\in \partial\calO^0,
      \\
      v(\zf)\geq1,~&\text{ on } ~\zf\in \partial\calO_R,\\
   v(\zf)\geq0, ~&\text{ on } ~\zf\in \partial\calO_{c/r}.   
\end{cases}
\end{align}
\item $v$ is a  viscosity solution of
  (\ref{dpe.crude}) if
  $v$ is both viscosity sub-solution and super-solution. 
\end{enumerate}
\end{definition}
\begin{remark}
  The definition of viscosity solutions is inextricably involved with min/max
  when the boundary conditions are given on derivatives. Consider
  $(p, X)\in \overline{\calJ}^{2, \pm}_{\calO} \varphi(\zf)$ for some
  $\zf \in \partial\calO^0$, where $\overline{\calJ}^{2, \pm}_{\calO}$ denote the closure of the
  second order superjet/subjet. Then there exists
  $(\zf_n, p_n, X_n) \in \calJ^{2, \pm}_{\calO}$ such that
  $(\zf_n, p_n, X_n) \to (\zf, p, X)$. However, in this case, we cannot guarantee
  that $\zf_n\in \partial\calO^0$ for any $n \in \dsN$. For more detailed discussion,
  readers can refer to Section 7 in \cite{crandall1992user}.
\end{remark}

Now, we are ready to state the main result of this paper.
\begin{theorem}[The Main Theorem]\label{thm.main}
  The value function $V$, defined at \cref{def.value}, is a unique viscosity
  solution of the HJB equation \cref{dpe.crude}.
\end{theorem}
The proof of the theorem is split into several steps, which will be provided in
the next sections.  
In summary, the first step is to define \textit{stochastic
  sub/super-solutions}. We continue to show that supremum (resp. infimum) of
\textit{stochastic sub-solutions} (resp. \textit{stochastic super-solutions}) is a viscosity
super-solution (resp. sub-solution). Then the main theorem can be concluded with
the help of the following comparison principle of the HJB equation, whose proof
is reported in the next section.
\begin{proposition}[Comparison Principle]\label{thm.comp}
Let $u$ and $v$ be a sub-solution and super-solution of (\ref{dpe.crude}),
respectively. Then $u \leq v$ in $\cl(\calO)$.
\end{proposition}
\section{Stochastic Perron's Method and Proofs}\label{sec-4}
This section contributes to the proof of Theorem \ref{thm.main} using stochastic
Perron's method, which helps us to avoid the lengthy and technical proof of
dynamic programming principle. To begin, we first need the concept of
\textit{random initial conditions} and \textit{exit times}.
\begin{definition}
 We call $(\tau, \xi)$  a \textit{random initial condition} if $\tau$ is a
 $\dsG$-stopping time valued in $\llbracket 0, \tau_D\rrbracket$, $\xi=(\xi^X, \xi^{Y^1},
\xi^{Y^2} )$ is a $\calG_\tau$-measurable random variable valued in $\cl(\calO) \cup \{\CST\}$, and $\xi = \CST$ if and only if
 $\tau = \tau_D$. We denote $\scrR$ as the set of all \textit{random initial
   conditions}. 
\end{definition}
\begin{definition}
The \textit{exit time} of $X^{\tau, \xi, \pi}$ from $\calO$, denoted by $\tau_E^{\tau,\xi,
  \pi}$, is defined by 
  $$\tau_E^{\tau, \xi, \pi} \coloneqq \inf\{ t\geq\tau \colon~X^{\tau, \xi, \pi}_t \notin\calO\}.$$
\end{definition}
\subsection{Stochastic Sub-solutions}
\label{sec:ssub}
This subsection first introduces the definition of \textit{stochastic sub-solutions} of
\cref{dpe.crude} and establishes the result that the stochastic envelope of
\textit{stochastic sub-solutions} is a viscosity super-solution of \cref{dpe.crude}. In a nutshell,
stochastic sub-solutions are functions that become $\dsG$-submartingales by
operating on $Z = (X, Y)$. The purpose of defining the stochastic sub-solutions
is to provide one direction of dynamic programming principle to some extent that
\begin{align}
  \inf_{\pi\in \scrA}\sup_{\dsQ\in \scrL}\dsE^\dsQ\bigg[V(Z^{\tau, \xi, \pi}_\rho)
  -\frac{1}{2\varepsilon}\int_{\tau}^{\rho}\Vert\theta_t \Vert^2\df t~\bigg|~ \calG_\tau\bigg] \geq  V(\xi),
  \label{one.direction}
\end{align}
for any \textit{random initial condition} $(\tau, \xi)$ and $\dsG$-stopping time
$\rho$ such that $\tau \leq \rho$. 
\begin{definition}[Stochastic sub-solutions]\label{def.sto.sub}
If $v \in \LSC_b(\cl(\calO))$ satisfies
\begin{enumerate}[label=(SB{\arabic*})]
\item  $v \leq 1$ on $\partial \calO_R$ and $v \leq0$ on  $\partial \calO_{c/r}$,
\item for any $(\tau, \xi)\in \mathscr{R}$, $\pi \in \scrA$, and
  $\dsG$-stopping time $\rho \in [ \tau, \tau_E^{\tau, \xi, \pi}]$,
  there exists a $\theta$-measure $\dsQ \in \scrL $ 
  such that
\begin{align}
\dsE^\dsQ\bigg[v(Z^{\tau, \xi, \pi}_\rho)
  -\frac{1}{2\varepsilon}\int_{\tau}^{\rho}e^{-\lambda^D s}\Vert\theta_t \Vert^2\df t~\bigg|~ \calG_\tau\bigg] \geq  v(\xi), \label{sb2}
\end{align}
where $v$ in (\ref{sb2}) is  understood as its extension to $\cl(\calO) \cup
\{\CST\}$ by allocating $v (\CST)=0$,
\end{enumerate}
then $v$ is called a stochastic sub-solution of \cref{dpe.crude}. In addition,
we denote by $\calV^-$ the class of all stochastic sub-solutions of
\cref{dpe.crude}.
\end{definition}
For the remaining of the paper, stochastic sub-solutions means stochastic
  sub-solutions of \cref{dpe.crude}. In addition, to understand the meaning of
the extension up to $\CST$, we can simply consider $\tau = \tau_D$ and derive that
$\tau=\tau_E=\tau_D=\rho$ and $\xi=\CST$. Then both sides in \cref{sb2} equal to zero and
the equation is trivially satisfied.  
\begin{remark}
  Note that we do not impose the oblique-type boundary condition $\calB$ arising
  from the high-watermark fees in the definition of stochastic
  sub-solutions. The Dirichlet boundary conditions are from the associated
  financial problems, namely the ruin probability minimization problem. Such
  boundary conditions are invariant given the underlying processes, i.e., the
  same Dirichlet boundary conditions are imposed regardless of the SDE for
  $Z=(X, Y)$.  However, the oblique-type boundary condition $\calB$ comes from
  the structure of the process, the running maximum of the process, as $\calF$
  does. Therefore, we can deal with $\calB$ and $\calF$ together in the same
  manner in applying SPM.  
  This in turn shows another advantage of
  stochastic Perron's method that is effective to handle control problem with
  high-watermark fee, especially with multiple hedge funds. Therefore, it is
  redundant to include the oblique-type boundary condition in
  \Cref{def.sto.sub}, which actually will make the argument more complicated
  because it is difficult to verify that $\calV^-$ is closed under the maximum
  operation with condition $\calB$.
\end{remark}
 Our first task is to find one stochastic
  sub-solution so that $\calV^-$ is not empty. One can think of \cref{sb2} as an upper-bound, in other words,
stochastic sub-solution can be found by considering a ``better
situation''. If there is no fee in reaching the high-watermark, the case is
clearly better for the investor. The minimal ruin probability in this
frictionless market was already studied by \cite{young2004optimal,bayraktar2007minimizing}, which  will turn out to be a
stochastic sub-solution in our case. Put
\begin{align} 
  \frU(x) \coloneqq&
  \begin{cases}
    \big(\frac{c-rx}{c-rR}\big)^\kappa, \quad &R\leq x\leq c/r,\\
  0, &c/r<x,
\end{cases}\nonumber\\
         \kappa \coloneqq
  & \frac{1}{2r}\big[(r+\lambda^D+R)+\sqrt{(r+\lambda^D+R)^2-4r\lambda^D}\big],\nonumber\\
  \Sigma\coloneqq &\frac{1}{2}\mu_\Delta^\top(\sigma \sigma^\top)^{-1}\mu_\Delta.\nonumber
\end{align}
Before proceeding, note that $\frU$ is a solution of the following
differential equation:
\begin{align}
  \begin{cases}
   &\lambda^D \frU(x) + \Sigma[\frU'(x)]^2/\frU''(x) +(c -rx)\frU'(x) = 0, \quad R<x<c/r,\\
  &\frU(R)=1, \quad \frU(c/r)=0.  
  \end{cases}
\end{align}
\begin{lemma}\label{lem.one.sub}
Let $\uup (x, y)\coloneqq \frU(x)$. Then $\uup \in \calV^-$.
\begin{proof}  
  It is obvious that $\uup$ is continuous and satisfies (SB1) in
  \cref{def.sto.sub}. To prove that $\uup$ is a stochastic sub-solution, let us
  consider an arbitrary \textit{random initial condition} $(\tau, \xi)$,
  $\pi \in \scrA$, and a $\dsG$-stopping time
  $\rho \in [\tau, \tau_E^{\tau, \xi, \pi}]$. Then we will show that (SB2) is satisfied with the
  reference measure $\dsP$. In other words, we choose
\begin{align}
\theta=0  \label{measure.choice}
\end{align}
in the representation of \cref{def.Q}. For the rest of this proof, we
omit the super-scripts $\tau, \xi, \pi$ for simplicity. Define a process
$(\Xo_t)_{t\geq0}$ given by $\Xo_\tau = X_\tau$, $\Xo_{\tau_D} \coloneqq \CST$, and
\begin{align}
  \df \Xo_t &= [r\Xo_t-c + \pi^\top_t\mu_\Delta]\df t +\pi^\top_t\sigma \df W_t,~~ \quad \text{for }~~ t < \tau_D. \nonumber
\end{align}
In other words, $\Xo$ is a process without high-watermark fees, thus $X
\leq \Xo$ on $\llbracket \tau, \tau_E \rrbracket$. As $\frU$ is non-increasing in $[R,
\infty)$,
\begin{align}
  \dsE[\uup(X_\rho, Y_\rho)\vert\calG_\tau] = \dsE[\1_{\rho < \tau_D}\frU(X_\rho)\vert\calG_\tau]
  \geq\dsE[\1_{\rho < \tau_D}\frU(\Xo_\rho)\vert\calG_\tau]=\dsE[\frU(\Xo_\rho)\vert\calG_\tau] \label{hihi}
\end{align}
Then, it suffices to show
$\dsE[\frU(\Xo_\rho)\vert\calG_\tau] \geq \frU(\Xo_\tau) (= \uup(\xi))$. We first consider the event
$U\coloneqq\{\Xo_\tau \in [R, c/r)\} \in \calG_\tau$ and let
$\nu \coloneqq \inf\{t \geq\tau \colon ~ \Xo_t \geq c/r\}$. On the event $U$, 
\begin{align}
  \dsE[\frU(\Xo_\rho)\vert\calG_\tau] \geq \dsE[\frU(\Xo_{\rho\wedge \nu})\vert\calG_\tau].\nonumber
\end{align}
In addition, applying It\^o's formula on the event $U$ yields
\begin{align}
  \frU (\Xo_{\rho\wedge \nu}) = &\frU (\Xo_\tau) + \int_{\tau}^{\rho\wedge \nu}\Big\{
  \frU'(\Xo_t)[(r\Xo_t-c) +\pi^\top_t\mu_\Delta] + \frU''(\Xo_t)\frac{1}{2}\Vert\sigma^\top\pi_t\Vert
  -\lambda^D \frU(\Xo_t) 
  \Big\} \df t\nonumber\\
  &+\int_{\tau}^{\rho\wedge \nu}\frU'(\Xo_t)\pi^\top_t\sigma  \df W_t
    -\int_{\tau}^{\rho\wedge \nu} \frU(\Xo_{t-})\df \calM^D_t   \nonumber
\end{align}
The $\df t$-integral term is non-negative. Moreover, $\frU$, $\frU'$, and $\pi$
are bounded, so the local martingales terms are martingales. Therefore, we have
\begin{align}
 \1_U\dsE[\frU (\Xo_{\rho\wedge \nu})\mid \calG_\tau] \geq \1_U\frU (\Xo_\tau).\label{hihi2}
\end{align}
On the other hand, on the event 
$U^c = \{\Xo_\tau \in [c/r, \infty) \cup\{\CST\}\}$, it clearly follows that
$\frU(\Xo_\tau)=0\leq \dsE[\frU(\Xo_\rho)\vert\calG_\tau]$. Therefore, thanks to
\cref{hihi}-\cref{hihi2}, we obtain
\begin{align}
 \dsE[\uup(X_\rho, Y_\rho)\vert\calG_\tau] &\geq \dsE[\frU(\Xo_\rho)\vert\calG_\tau]  =
   \dsE[\1_{U}\frU(\Xo_\rho) + \1_{U^c}\frU(\Xo_\rho)\vert\calG_\tau] \nonumber\\
  &\geq \1_{U}\frU(\Xo_\tau) = \frU(\Xo_\tau)\nonumber\\
  &=\uup(\xi).
\end{align}
Thus by \cref{measure.choice}, $\uup$ satisfies (SB2).
\end{proof}
\end{lemma}
To show the stochastic envelope of stochastic sub-solutions is a viscosity
super-solution, we first show $\calV^-$ is closed under maximum operation. 
\begin{lemma}\label{lem.max.sub}
If $v^1, v^2 \in \calV^-$, then $ v^1 \vee v^2 \in \calV^-$.
\begin{proof}
  It is easy to check that $ v^1 \vee v^2\in \LSC_b(\clco)$ and $v^1\vee v^2$ satisfies
  (SB1) in \cref{def.sto.sub}. Let $(\tau, \xi) \in \scrR$, $\pi \in \scrA$, $\rho$ be a
  $\dsG$-stopping time valued in interval
  $[ \tau, \tau^{\tau, \xi, \pi}_E]$. Because $v^1$ and $v^2$ are
  stochastic sub-solutions, there exist $\dsQ^1$ and $\dsQ^2$
  satisfying (SB2). We denote by $\theta^i$, $i \in \{1, 2\}$, the processes that
  generate $\dsQ^i$.  To find the measure satisfying (SB2) for $v^1 \vee v^2$, we
  define $\frA \coloneqq \{v^1(\xi) > v^2(\xi)\} \in \calG_\tau$, 
  $ \theta \coloneqq \1_{\llbracket \tau, \infty\rrbracket}[\1_{\frA}\theta^1 + \1_{\frA^c}\theta^2]$,
  and let $\dsQ$ denote the measure generated by $\theta$, i.e.,
  on the stochastic interval $\llbracket \tau, \infty\rrbracket$,
\begin{align}
  \frac{\df \dsQ}{\df \dsP}\bigg|_{\calG_\cdot} =
  \1_\frA\frac{\df \dsQ^1}{\df \dsP}\bigg|_{\calG_\cdot}+
   1_{\frA^c}\frac{\df \dsQ^2}{\df \dsP}\bigg|_{\calG_\cdot} .\nonumber
\end{align}
 Then as $v^1$ is a stochastic sub-solution and $\frA\in \calG_\tau$, we have
\begin{align}
  \1_{\frA}v^1(\xi) \leq
  & \1_\frA\dsE^{\dsQ^1}\bigg[v^1(Z^{\tau, \xi, \pi}_\rho) -\frac{1}{2\varepsilon}\int_{\tau}^{\rho}e^{-\lambda^D
    s}\Vert\theta^1_s \Vert^2 \df s ~\bigg|~ \calG_\tau\bigg]\nonumber\\
  =&
    \dsE\bigg[\1_\frA\frac{\df \dsQ^1}{\df \dsP}\bigg|_{\calG_{\rho}}
    \bigg\{v^1(Z^{\tau, \xi, \pi}_\rho) -\frac{1}{2\varepsilon}\int_{\tau}^{\rho}e^{-\lambda^D
     s}\Vert\theta^1_s \Vert^2 \df s\bigg\} ~\bigg|~ \calG_\tau\bigg]\nonumber\\
  =& \dsE\bigg[\1_\frA\frac{\df \dsQ}{\df \dsP}\bigg|_{\calG_{\rho}}
    \bigg\{v^1(Z^{\tau, \xi, \pi}_\rho) -\frac{1}{2\varepsilon}\int_{\tau}^{\rho}e^{-\lambda^D
    s}\Vert\theta_s \Vert^2 \df s\bigg\} ~\bigg|~ \calG_\tau\bigg]\nonumber\\
  \leq&\1_\frA\dsE^{\dsQ}\bigg[ (v^1\vee v^2)(Z^{\tau, \xi, \pi}_\rho) -\frac{1}{2\varepsilon}\int_{\tau}^{\rho}e^{-\lambda^D
    s}\Vert\theta_s \Vert^2 \df s ~\bigg|~ \calG_\tau\bigg].\label{max.eq1}
\end{align}
The second equality above is obtained by \cref{def.theta} and boundness of
$v^1$.  Similarly, we obtain
\begin{align}
  \1_{\frA^c}v^2(\xi) \leq \1_{\frA^c}\dsE^{\dsQ}\bigg[ (v^1\vee v^2)(Z^{\tau, \xi, \pi}_\rho)
  -\frac{1}{2\varepsilon}\int_{\tau}^{\rho}e^{-\lambda^D
  s}\Vert\theta_s \Vert^2 \df s ~\bigg|~ \calG_\tau\bigg].\label{max.eq2}
\end{align}
Combining \cref{max.eq1} and \cref{max.eq2}, we have
\begin{align}
(v^1\vee v^2)(\xi)\leq    \dsE^{\dsQ}\bigg[ (v^1\vee v^2)(Z^{\tau, \xi, \pi}_\rho)
  -\frac{1}{2\varepsilon}\int_{\tau}^{\rho}e^{-\lambda^D s}\Vert\theta_s \Vert^2 \df s ~\bigg|~ \calG_\tau\bigg].\nonumber
\end{align}
Thus, $(v^1 \vee v^2)$ satisfies (SB2) with $\dsQ$. 
\end{proof}
\end{lemma}
In the next theorem, we will use \cref{lem.max.sub} to construct a ``bump''
function to argue by contradiction.
\begin{theorem}\label{thm.v-}  
  The lower stochastic envelope of $\calV^-$,
  \begin{align}
    v^- \coloneqq \sup_{v \in \calV^-} v, \label{sup.sto.sub}
  \end{align}
is a viscosity super-solution of \cref{dpe.crude}.
\begin{proof}
\Cref{lem.one.sub} already asserts that $v^- \geq \uup$. Therefore, we have $v^- \geq 1$ on $\calO_R$
and $v^- \geq0$ on $\calO_{c/r}$. It remains to show that for this $v^-$ and any test function $\varphi$ such that $ \zf \in \calO$ is a
minimum point of $v^- - \varphi$ at zero, we have
\begin{align}
\begin{cases}
 \calF[\varphi](\zf)\geq0, ~&\text{ on }~
     \zf \in \calO^+, \\
\max\big\{\calF[\varphi](\zf), ~\calB^*[\varphi](\zf)\big\} \geq0,
 ~&\text{ on } ~\zf\in \partial\calO^0.
\end{cases}\nonumber
\end{align}
We first show the claim above holds on the boundary part
$\partial\calO^0_1 \cap \partial\calO^0_2$.

Let us consider the region $B_a(\zf_0)$ of a
ball with center $\zf_0 \in \cl(\calO)$ and the radius $a$ intersecting with
$\calO$ that
\begin{align}
 B_a(\zf_0)\coloneqq \{\zf \in \cl(\calO)\colon~\Vert\zf -\zf_0 \Vert < a\}. \nonumber
\end{align}
To argue by contradiction, we suppose that there exist 
$\zf_0=(x_0, 0, 0)\in \partial\calO^0_1 \cap \partial\calO^0_2$ and some
$\varphi\in C^2(\calO)$ such that $v^- -\varphi$ attains its strict minimum of
zero at $\zf_0$ and
\begin{align}\label{calFB}
  \max\big\{\calF[\varphi](\zf_0), ~\calB^1[\varphi](\zf_0), ~\calB^2[\varphi](\zf_0)\big\} <0.
\end{align}
Therefore it follows that there exists a constant $\theta^\varphi \in \calL$ such that
\begin{align}
 \lambda^D\varphi(\zf_0) -(rx_0-c)\varphi_x(\zf_0)-\inf\limits_{\pi\in \calK}
             \calA^{\pi, \theta^{\varphi}}[\varphi](\zf_0) <0 .
\end{align}
Using $\varphi$, we will construct a bump function that still is in $\calV^-$,
in which it contradicts to \cref{sup.sto.sub}.  By continuity of $\calF$ and
$\calB^i$, $i \in \{1,2\}$, we can choose a small ball
$B_{2a}(\zf_0)$, $a>0$,  such that for any $\zf \in \cl(B_{2a}(\zf_0))$,
\begin{align}\label{calFB2}
  \max\Big\{\lambda^D\varphi(\zf) -(rx-c)\varphi_x(\zf)-\inf\limits_{\pi\in \calK}
             \calA^{\pi, \theta^{\varphi}}[\varphi](\zf)
  <0,~\calB^1[\varphi](\zf),~\calB^2[\varphi](\zf)
  \Big\} <0.
\end{align}

 As $v^- -
\varphi$ is l.s.c and $\cl(B_{2a}(\zf_0)) \setminus B_{a}(\zf_0)$ is compact, there
exists $\delta >0$ satisfying
\begin{align}
  v^- - \varphi \geq \delta, \quad \text{on}~~\cl(B_{2a}(\zf_0)) \setminus B_a(\zf_0). \nonumber
\end{align}
As a result of Proposition 4.1 in \cite{bayraktar2012stochastic} and
\Cref{lem.max.sub}, we can choose a non-decreasing sequence $\{v_n\}  \subseteq \calV^-$
such that $v_n\nearrow v^-$. By Lemma 2.4 in \cite{bayraktar2014stochastic}, we can pick
$v\coloneqq v_N$ such that
\begin{align}
  v - \varphi \geq \delta/2~~ \text{on}~~\cl(B_{2a}(\zf_0)) \setminus B_a(\zf_0).\nonumber
\end{align}
Then we further choose $0< \eta < \delta/2$ small enough such that 
$\varphi^\eta \coloneqq \varphi +\eta$ satisfies 
\begin{align}\label{eq.calFB}
\max\Big\{  \lambda^D\varphi^\eta(\zf) -(rx-c)\varphi^\eta_x(\zf)-\inf\limits_{\pi\in \calK}
             \calA^{\pi, \theta^{\varphi}}[\varphi^\eta](\zf),
  ~\calB^1[\varphi^\eta](\zf),~\calB^2[\varphi^\eta](\zf)
  \Big\}<0,
\end{align}
on $\cl(B_{2a}(\zf_0))$.
By this construction, we have
\begin{align}
  \varphi^\eta \leq& \varphi + \delta/2 \leq v ~~\text{on}~~ \cl(B_{2a}(\zf_0)) \setminus B_a(\zf_0),\\
  \varphi^\eta(\zf_0) =& \varphi(\zf_0) + \eta=v^-(\zf_0) + \eta > v^-(\zf_0). \label{comp.cont}
\end{align}
Let us define
\begin{align}
  v^\eta \coloneqq
  \begin{cases}
 v \vee \varphi^\eta,  ~  &\cl(B_{2a}(\zf_0)),\\
v, &\text{otherwise}.
  \end{cases}\nonumber
\end{align}
Then we will show that $v^\eta \in \calV^-$ and this is a contradiction by
\cref{sup.sto.sub} and \cref{comp.cont}.

To this end, we consider an arbitrary $(\tau, \xi) \in \scrR$, $\pi \in \scrA$, and a
$\dsG$-stopping time $\rho \in \llbracket \tau, \tau_E^{\tau, \xi, \pi} \rrbracket$. Our goal is to find a probability
measure satisfying (SB2) for $v^\eta$.  As $v$ is a  stochastic
  sub-solution, for any strategy $\pi$ we can find
$(\theta^{v, \pi}_t)_{t\geq0}$ producing a probability measure
$\dsQ^{v, \pi} \in \scrL$ that satisfies (SB2) for $v$. Define
\begin{align}
  \Gamma \coloneqq \big\{
  \xi \in B_a(\zf_0) \text{ and } v(\xi) < \varphi ^\eta(\xi)
  \big\}\in \calG_\tau,\nonumber
\end{align}
and let $\tau_{a}$ (resp. $\xi_{a}$) denote the exit time (resp. exit
position) of the ball $B_a(\zf_0)$, i.e.,
\begin{align}
  \tau_{a}\coloneqq &\inf\{t \in [\tau, \tau_E^{\tau, \xi, \pi}] \colon~
  Z^{\tau, \xi, \pi}_t\notin B_a(\zf_0)\},\nonumber\\
  \xi_{a}\coloneqq &Z_{\tau_a}^{\tau, \xi, \pi}.  \nonumber
\end{align}
By $(\theta^{v, \pi}_t)_{t\geq0}$ and $\theta^{\varphi}$  in \cref{eq.calFB}, define  $(\jt_t)_{t\geq0}$ as
\begin{align}
  \jt^\pi_t \coloneqq \1_{t\geq\tau}(\theta^{\varphi} \1_{\Gamma} + \theta^{v, \pi}_t\1_{\Gamma^c})\nonumber
\end{align}
Note that $\xi_{a}\in \partial B_a(\zf_0) \cup \{\CST\}$  and 
$(\tau_{a}, \xi_{a}) \in \scrR$. Therefore, for $(\tau_a, \xi_a)$ and $\pi\in \scrA$, there
exists $\theta^{v, a, \pi}$ producing $\dsQ^{v, a, \pi}$ given by \cref{def.Q} that satisfies (SB2)
for $v$. Then define 
\begin{align}
  \theta^\pi \coloneqq \1_{\llbracket 0, \tau_a\rrbracket}\jt^\pi
  +  \1_{\rrbracket \tau_a, \infty \rrbracket}\theta^{v, a, \pi}\label{st.cont},
\end{align}
and  $\dsQ^\pi$ be the measure by $\theta^\pi$.
Then for any $\pi \in \scrR$, we show that $\dsQ^\pi$ is the measure for  $v^\eta$ to
satisfy (SB2) from which we obtain the contradiction.

In particular, we can obtain a contradiction from the place where the measure by
$\theta^{\varphi, \pi}$ is taken. It\^o's formula on the event $\Gamma$ yields 
\begin{align}
  \varphi^\eta(Z^{\tau, \xi, \pi}_{\rho \wedge \tau_a}) -\varphi^\eta(Z^{\tau, \xi, \pi}_{\tau})
  =&\int_{\tau}^{\rho \wedge \tau_{a}}\big[\calA^{\pi, \theta^\pi}[\varphi^\eta] + \frac{1}{2\varepsilon}\Vert\theta^\pi_t\Vert^2 - \lambda^D\varphi^\eta
     + (rX_t^{\tau, \xi, \pi}-c)\varphi^\eta_x\big]
     (Z^{\tau, \xi, \pi}_{t}) \df t \nonumber\\
  &- \SUM_{i=1,2}\int_{\tau}^{\rho \wedge \tau_a}\calB^i[\varphi^\eta](Z^{\tau, \xi, \pi}_t)\df M^i_t\nonumber\\
  &-\int_{\tau}^{\rho\wedge\tau_{a}}\varphi^\eta(Z^{\tau, \xi, \pi}_{t-})\df \calM^D_t
 \nonumber\\
   &+\int_{\tau}^{\rho\wedge\tau_{a}}\big[\varphi^\eta_x(Z^{\tau, \xi, \pi}_{t})\pi^\top_t\sigma - \nabla_y\varphi^\eta
     (Z^{\tau, \xi, \pi}_{t})^\top\diag(\pi_t)\sigma^B_\Delta\big] \df W^{\dsQ^\pi}_t.
    \label{ito.perron1}
\end{align} 
On the compact set $\cl(B_{a}(\zf_0))$, $\varphi^\eta$ and $\nabla \varphi^\eta$ are
bounded. Therefore, $\varphi^\eta(Z^{\tau, \xi, \pi})$ and $\nabla\varphi^\eta(Z^{\tau, \xi, \pi})$ are
bounded on
 $\llbracket \tau, \rho\wedge\tau_{a} \rrbracket$. Moreover, $\pi$ is valued in the compact set
 $\calK$. Therefore, the last two terms in
\cref{ito.perron1} are $\dsG$-martingales. Then by \cref{calFB}, we have
\begin{align}
  \dsE^{\dsQ^\pi}\big[\1_{\Gamma}v^\eta\big(Z^{\tau, \xi, \pi}_{\rho\wedge \tau_{a}}\big)\big\vert\calG_\tau\big]
  &\geq\dsE^{\dsQ^\pi}
     \bigg[\1_{\Gamma}\Big\{\varphi^\eta\big(Z^{\tau, \xi, \pi}_{\rho\wedge \tau_{a}}\big)
     -\frac{1}{2\varepsilon}\int_{\tau}^{\tau_a\wedge \rho}\Vert \theta_t^\pi\Vert^2\df t\Big\}\bigg\vert\calG_\tau\bigg]
     \nonumber\\
  &\geq\1_{\Gamma}\varphi^\eta\big(Z^{\tau, \xi, \pi}_{\tau}\big) =\1_\Gamma\varphi^\eta(\xi)\nonumber\\
  &= \1_\Gamma v^\eta(\xi).\nonumber
\end{align}
Note that at the last equality, we do not exclude the case that $\tau = \tau^D$, i.e.,
$\xi = \CST$.  Recall that on $\Gamma^c$, we have $v(\xi) = v^\eta(\xi)$ and $\theta^\pi = \theta^{v, \pi}$ which is the
$(\tau, \xi)$-optimal control of $v$. Let $\dsQ^{v, \pi}$ denote the $\theta^{v,
  \pi}$-measure. By (SB2), it follows that
\begin{align}
  \1_{\Gamma^c} v^\eta(\xi)= \1_{\Gamma^c} v(\xi) &\leq
  \dsE^{\dsQ^{v, \pi}}\bigg[\1_{\Gamma^c}\Big\{v\big(Z^{\tau, \xi, \pi}_{\rho\wedge
    \tau_{a}}\big)-\frac{1}{2\varepsilon}\int_{\tau}^{\tau_a\wedge \rho}\Vert \theta_t^{v, \pi}\Vert^2\df
    t\Big\}\bigg\vert\calG_\tau\bigg]
    \nonumber\\
  &\leq \dsE^{\dsQ^\pi}\bigg[\1_{\Gamma^c}\Big\{v^\eta\big(Z^{\tau, \xi, \pi}_{\rho\wedge \tau_{a}}\big)
     -\frac{1}{2\varepsilon}\int_{\tau}^{\tau_a\wedge \rho}\Vert \theta_t^{\pi}\Vert^2\df t\Big\}\bigg\vert\calG_\tau\bigg].\nonumber
\end{align}
Hence, we obtain that
\begin{align}
  v^\eta(\xi) \leq \dsE^{\dsQ^\pi}\bigg[v^\eta\big(Z^{\tau, \xi, \pi}_{\rho\wedge \tau_{a}}\big)
  -\frac{1}{2\varepsilon}\int_{\tau}^{\tau_a\wedge \rho}\Vert \theta_t^{\pi}\Vert^2\df t\bigg|\calG_\tau\bigg].
  \label{upto.min}
\end{align}
Now, to replace $ \rho\wedge \tau_a$  with $\rho$ in \cref{upto.min}, we first consider the
event $\Lambda \coloneqq\{\rho > \tau_a\}\in \calG_{\tau_a \wedge \rho}$.  Since
$v = v^\eta$ at $\partial B_a(\zf_0)$ and on
$ \rrbracket \tau_a, \rho\rrbracket \cap (\Lambda\times\dsR_+) $, we have
$\theta^\pi = \theta^{v, a, \pi}$. Then denoting by $\dsQ^{v, a, \pi}$ the $\theta^{v, a,
  \pi}$-measure, 
\begin{align}
  \1_\Lambda v^\eta(\xi_a) = \1_\Lambda v(\xi_a) &\leq
   \dsE^{\dsQ^{v, a, \pi}} \bigg[\1_\Lambda \Big\{v(Z^{\tau, \xi, \pi}_\rho)
    -\frac{1}{2\varepsilon}\int_{\tau_a}^{\rho}\Vert \theta_t^{v, a, \pi}\Vert^2\df t\Big\}\bigg|\calG_{\tau_a}\bigg]\nonumber\\
  &\leq \dsE^{\dsQ^\pi}\bigg[\1_\Lambda \big\{v^\eta(Z^{\tau, \xi, \pi}_\rho)
     -\frac{1}{2\varepsilon}\int_{\tau_a}^{\rho}\Vert \theta_t^{\pi}\Vert^2\df t\Big\}\bigg|\calG_{\tau_a}\bigg]. \label{on.lambda}
\end{align}
Moreover, by (\ref{upto.min}) together with (\ref{on.lambda}), we can get
\begin{align}
  v^\eta(\xi) &\leq \dsE^{\dsQ^\pi}\bigg[v^\eta(Z^{\tau, \xi, \pi}_{\rho \wedge \tau_a})
            -\frac{1}{2\varepsilon}\int_{\tau}^{\tau_a\wedge \rho}\Vert \theta_t^{\pi}\Vert^2\df t\bigg|\calG_{\tau}\bigg] \nonumber\\
  &=\dsE^{\dsQ^\pi}\bigg[\1_{\Lambda^c} \Big\{v^\eta(Z^{\tau, \xi, \pi}_{\rho})
     -\frac{1}{2\varepsilon}\int_{\tau}^{\rho}\Vert \theta_t^{\pi}\Vert^2\df t\Big\}
     + \1_\Lambda \Big\{v^\eta(\xi_a)-\frac{1}{2\varepsilon}\int_{\tau}^{\tau_a}\Vert \theta_t^{\pi}\Vert^2\df t
     \Big\}\bigg|\calG_{\tau}\bigg].\label{super.split}
\end{align}
By \cref{on.lambda}, we have
\begin{align}
\dsE^{\dsQ^\pi}  \bigg[\1_\Lambda \Big\{v^\eta(\xi_a)-\frac{1}{2\varepsilon}\int_{\tau}^{\tau_a}\Vert \theta_t^{\pi}\Vert^2\df t
     \Big\}\bigg|\calG_{\tau}\bigg]
  &= 
    \dsE^{\dsQ^\pi}  \bigg[\dsE^{\dsQ^\pi}\bigg[\1_\Lambda
    \Big\{v^\eta(\xi_a)-\frac{1}{2\varepsilon}\int_{\tau}^{\tau_a}\Vert \theta_t^{\pi}\Vert^2\df t
    \Big\}\bigg|\calG_{\tau_a}\bigg]\bigg|\calG_{\tau}\bigg] \nonumber\\
  &\leq \dsE^{\dsQ^\pi}  \bigg[\1_\Lambda \Big\{v^\eta(Z_\rho^{\tau, \xi, \pi})-\frac{1}{2\varepsilon}\int_{\tau}^{\rho}\Vert \theta_t^{\pi}\Vert^2\df t
    \Big\}\bigg|\calG_{\tau}\bigg]. \label{on.lambda2}
\end{align}
Therefore, in view of \cref{super.split} and \cref{on.lambda2}, we deduce that
$v^\eta \in \calV^-$, which clearly contradicts (\ref{sup.sto.sub}). Hence, it
follows that $v^-$ is a viscosity super-solution of \cref{dpe.crude} at
$\zf_0 \in \partial\calO_1\cap\partial\calO_2$.

We can deal with points in other regions
$\zf_0 \notin \partial\calO_1\cap\partial\calO_2$ in similar ways. To be more precise, for
$\zf_0 \in \calO^+$ (resp.  $\zf_0 \in \partial\calO_i$, $i \in \{1, 2\}$), we suppose that
there exist a function $\varphi\in C^2(\calO)$ such that $v^- -\varphi$ attains
its strict minimum of zero at $\zf_0$ and
\begin{eqnarray}
  &\calF[\varphi](\zf_0) <0\nonumber\\
  &\big( \text{resp}. ~\max\{\calF[\varphi](\zf_0),
  ~\calB^i[\varphi](\zf_0)\} <0 \big).\nonumber
\end{eqnarray}
Then, by employing similar contradiction arguments, we can conclude that $v^-$
is indeed a viscosity super-solution of \cref{dpe.crude}.
\end{proof}
\end{theorem}

\subsection{Stochastic Super-solutions}
\label{sec:ssuper}
Roughly speaking, \textit{stochastic super-solutions} can be defined to facilitate the derivation of the other direction of DPP as
\begin{align}
\inf_{\pi\in \scrA}\sup_{\dsQ\in \scrL}\dsE^\dsQ\bigg[V(Z^{\tau, \xi, \pi}_\rho)
  -\frac{1}{2\varepsilon}\int_{\tau}^{\rho}e^{-\lambda^D s}\Vert\theta_s \Vert^2\df s~\bigg|~ \calG_\tau\bigg] \leq  V(\xi).\nonumber
\end{align}
Note that the item (SP2) in the next definition is precisely motivated by the inequality above.

\begin{definition}[Stochastic super-solutions]\label{def.sto.super}
  If $v \in \USC_b(\clco)$ satisfies
\begin{enumerate}[label=(SP{\arabic*})]
\item  $v \geq 1$ on $\partial \calO_R$ and $v \geq0$ on  $\partial \calO_{c/r}$,
\item for any random initial condition $(\tau, \xi)$, there exists
  $\pi \in \scrA$ such that for any $\dsG$-stopping time
  $\rho \in [\tau, \tau_E^{\tau, \xi, \pi}]$ and $\dsQ\in \scrL$,
\begin{align}
  \dsE^\dsQ\bigg[v(Z^{\tau, \xi, \pi}_\rho)
  -\frac{1}{2\varepsilon}\int_{\tau}^{\rho}e^{-\lambda^D s}\Vert\theta_s \Vert^2\df s~\bigg|~ \calG_\tau\bigg] \leq  v(\xi), \label{sP2}
\end{align}
where $v$ in (\ref{sb2}) is  understood as its extension to $\cl(\calO) \cup
\{\CST\}$ by allocating $v (\CST)=0$,
\end{enumerate}
then $v$ is called a stochastic super-solution of \cref{dpe.crude}. In addition,
we let $\calV^+$ denote the class of all stochastic super-solutions of
\cref{dpe.crude}.
\end{definition}
We can find a stochastic super-solution by considering a ``worse
scenario''. Consider a situation that the investor does not invest in the hedge funds, i.e., $\pi=0$. Then, the investor's wealth follows
$ \df X_t = [rX_t-c]\df t,~X_0=x$.
We thus, can  obtain that
\begin{align}
  \frp(x)\coloneqq \dsP(\tau^{x, y, 0}_R < \tau_D) = \Big(\frac{c-rx}{c-rR}\Big)^{\frac{\lambda^D}{r}}. \nonumber
\end{align}
\begin{lemma}\label{lem.one.sup} 
Let $\oup (x, y)\coloneqq \frp(x)$. Then $\oup \in \calV^+$.
\begin{proof}
  It is obvious that $\oup \in \USC_b(\clco)$ and satisfies (SP1).  Let
  $(\tau, \xi)$ be a \textit{random initial condition} and we choose $\pi =0$ for the
  strategy.  Thus, for $\tau < \tau_D$,
\begin{align}
  \df X^{\tau, \xi, \pi}_t = [rX^{\tau, \xi, \pi}_t  - c]\df t.\nonumber
\end{align}
Consider $\rho \in [ \tau, \tau_E^{\tau, \xi, \pi}]$
as a $\dsG$-stopping time. In the rest of the
proof,  we suppress the superscripts $\tau, \xi, \pi$. By It\^o's formula, we have
\begin{align}
  \frp(X_\rho) - \frp(X_\tau)=
  &\int_\tau^{\rho}\Big\{\frp'(X_t)[rX_t-c] - \lambda^D\frp(X_t)\Big\} -\int_{\tau}^{\rho}\frp(X_{s-})\df
    \calM^D_s
    \nonumber\\
  =&-\int_{\tau}^{\rho}\frp(X_{s-})\df
    \calM^D_s\nonumber
\end{align}
As for any equivalent probability measure $\dsQ$ given by \cref{def.Q},  $\calM^D$ is
$(\dsQ, \dsG)$-martingale, it follows that 
$ \dsE^{\dsQ}[\frp(X_\rho)|\calG_\tau] = \frp(X_\tau) $ for any $\dsQ\in \scrL$.
Therefore, for any $\theta$-measure $\dsQ\in \scrL$,
\begin{align}
  \dsE^{\dsQ}\bigg[\oup(Z_\rho)-\frac{1}{2\varepsilon}\int_{\tau}^{\rho}\Vert\theta_t\Vert^2 \df
  t\bigg|\calG_\tau\bigg]
  & = \dsE^{\dsQ}\bigg[\frp(Z_\rho)-\frac{1}{2\varepsilon}\int_{\tau}^{\rho}\Vert\theta_t\Vert^2 \df
    t\bigg|\calG_\tau\bigg] \nonumber\\
  &\leq\dsE^{\dsQ}[\frp(X_\rho)|\calG_\tau] = \frp(X_\tau)=\oup(\xi). \nonumber
\end{align}
Therefore, $\oup$ satisfies (SP2), and we can deduce that $\oup\in \calV^+$. 
\end{proof}
\end{lemma}
As in the previous section, we need to show $\calV^+$ is stable under minimum
operation. The proof follows closely the argument to prove
\Cref{lem.max.sub}, so we omit it.
\begin{lemma}\label{lem.min.super}
If $v^1, v^2 \in \calV^+$, then $ v^1 \wedge v^2 \in \calV^+$.
\end{lemma}
Then \cref{lem.min.super} will be used to construct a  bump function in the
following theorem.
\begin{theorem}\label{thm.v+}
The lower stochastic envelope of $\calV^+$,
\begin{align}
  v^+ \coloneqq \inf_{v \in \calV^+} v, \label{inf.sto.sup} 
\end{align}
is a viscosity sub-solution of \cref{dpe.crude}
\begin{proof}
By \Cref{lem.one.sup}, $v^+ \leq \oup$. Therefore, we have $v^+ \leq 1$ on $\calO_R$
and $v^+ \leq$ on $\calO_{c/r}$. As in the proof of \cref{thm.v-}, it is sufficient to verify
the sub-solution property of $v^+$ only on the boundary part $\partial\calO^0_1 \cap \partial\calO^0_2$. 
Using the same notation of balls that intersect $\calO$, we again will prove by contradiction. Suppose that there exist
$\zf_0=(x_0, 0, 0)\in \partial\calO^0_1 \cap \partial\calO^0_2$ and
$\varphi\in C^2(\calO)$ such that $v^- -\varphi$ attains its strict maximum of
zero at $\zf_0$ and
\begin{align}\label{calFB.s}  
  \min\big\{\calF[\varphi](\zf_0),
  ~\calB^1[\varphi](\zf_0), ~\calB^2[\varphi](\zf_0)\big\} >0.
\end{align}
Again, as in the construction of a bump function in \cref{thm.v-}, we can choose constants
$\pi^\varphi\in \calK$, $\eta>0, a>0$,  and a stochastic super-solution $v \in \calV^+$ such that
\begin{align}\label{calFB.sup}
\begin{cases}  
  \varphi^\eta=\varphi + \eta \geq v,\quad &\text{on}~~ \cl(B_{2a}(\zf_0)) \setminus B_a(\zf_0),\\
  \lambda^D\varphi^\eta -(rx-c)\varphi^\eta_x- \sup\limits_{\theta\in \calL}\calA^
  {\pi^\varphi, \theta}[\varphi^\eta] >0,\quad&\text{on}~~  \cl(B_{2a}(\zf_0)),\\
\min\big\{\calB^1[\varphi^\eta], ~\calB^2[\varphi^\eta]\big\} >0,
 \quad &\text{on}~~  \cl(B_{2a}(\zf_0)),\\
 \varphi^\eta(\zf_0) < v^-(\zf_0), 
\end{cases}  
\end{align}
and we define
\begin{align}\label{sup.varphi} 
  v^\eta \coloneqq
  \begin{cases}
 v \wedge \varphi^\eta,  ~  &\cl(B_{2a}(\zf_0)),\\
v, &\text{otherwise}.
  \end{cases} 
\end{align} 
Then we will show that $v^\eta \in \calV^+$. To show that $v^\eta$ satisfies (SP2), let
$(\tau, \xi)\in \scrR$. Since $v \in \calV^+$, we can choose $(\pi^v_t)_{t\geq0}$ for $v$ to satisfy
(SP2). Then with $\pi^\varphi$ in \cref{calFB.sup}, we define  $\tilde{\pi}_{t\geq0}$
as
\begin{align}
  \tilde{\pi}_t \coloneqq
\1_{t\geq\tau}(\pi^\varphi \1_{\Gamma} + \pi^v_t\1_{\Gamma^c}). \nonumber
\end{align}
Let us denote 
\begin{align}
  \Gamma \coloneqq \big\{\xi \in B_a(\zf_0) \text{ and } v(\xi) <  \varphi^\eta(\xi)\big\},\nonumber
\end{align}
and  let $\tau_{a}$ (resp. $\xi_{a}$) denote the exit time (resp. exit position) of
the ball $B_{a}(\zf_0)$. Since $(\tau_a, \xi_a)\in \scrR$ and $v \in \calV^+$, we can
choose  $\pi^{v, a} \in\scrA$ such that for any $\dsQ\in \scrL$ and $\dsG$-stopping time
valued in $\llbracket \tau_a, \tau_E^{\tau, \xi, \pi^{v, a}}\rrbracket$, $v$ satisfies (SP2).  Finally, we let
\begin{align}
  \pi \coloneqq \1_{\llbracket 0, \tau_a\rrbracket}\tilde{\pi}
  +  \1_{\rrbracket \tau_a, \infty \rrbracket}\pi^{v, a}\label{st.sup.cont}.
\end{align}
We will show that $v^\eta$, with $\pi$, satisfies (SP2).  
Consider  an arbitrary $\dsG$-stopping time $\rho \in [\tau, \tau_E^{\tau, \xi, \pi}]$ and
$\theta$-measure $\dsQ\in \scrL$. 
Applying \ito's formula on the event $\Gamma$ yields, for any $\theta$-measure $\dsQ$,
\begin{align}
  \varphi^\eta(Z^{\tau, \xi, \pi}_{\rho \wedge \tau_a}) -\varphi^\eta(Z^{\tau, \xi, \pi}_{\tau})
  =&\int_{\tau}^{\rho \wedge \tau_{a}}\big[\calA^{\pi, \theta}[\varphi^\eta] + \frac{1}{2\varepsilon}\Vert\theta_t\Vert^2 - \lambda^D\varphi^\eta
     + (rX_t^{\tau, \xi, \pi}-c)\varphi^\eta_x\big]
     (Z^{\tau, \xi, \pi}_{t}) \df t \nonumber\\
  &- \SUM_{i=1,2}\int_{\tau}^{\rho \wedge \tau_a}\calB^i[\varphi^\eta](Z^{\tau, \xi, \pi}_t)\df M^i_t\nonumber\\
  &-\int_{\tau}^{\rho\wedge\tau_{a}}\varphi^\eta(Z^{\tau, \xi, \pi}_{t-})\df \calM^D_t
 \nonumber\\
  &+\int_{\tau}^{\rho\wedge\tau_{a}}\big[\varphi^\eta_x(Z^{\tau, \xi, \pi}_{t})\pi^\top_t\sigma - \nabla_y\varphi^\eta(Z^{\tau, \xi, \pi}_{t})^\top\diag(\pi_t)\sigma^B_\Delta\big] \df W^{\dsQ}_t.
    \label{ito.perron2}
\end{align} 
Therefore, by \cref{calFB.sup} and \cref{sup.varphi}, we have  
\begin{align}
  \dsE^{\dsQ}\bigg[\1_{\Gamma}\Big\{
  v^\eta\big(Z^{\tau, \xi, \pi}_{\rho\wedge \tau_{a}}\big)
  -\frac{1}{2\varepsilon}\int_{\tau}^{\tau_a\wedge \rho}\Vert \theta_t\Vert^2\df t\Big\}\bigg\vert\calG_\tau\bigg]
  &\leq\dsE^{\dsQ}
     \bigg[\1_{\Gamma}\Big\{\varphi^\eta\big(Z^{\tau, \xi, \pi}_{\rho\wedge \tau_{a}}\big)
     -\frac{1}{2\varepsilon}\int_{\tau}^{\tau_a\wedge \rho}\Vert \theta_t\Vert^2\df t\Big\}\bigg\vert\calG_\tau\bigg]
     \nonumber\\
  &\leq\1_{\Gamma}\varphi^\eta\big(Z^{\tau, \xi, \pi}_{\tau}\big) =\1_\Gamma\varphi^\eta(\xi)\nonumber\\
  &= \1_\Gamma v^\eta(\xi).\nonumber
\end{align}
Recall that on $\Gamma^c$, we have $v(\xi) = v^\eta(\xi)$ and $\pi =\pi^v$.
Since $v$ is a stochastic super-solution by its construction, we have
\begin{align}
  \1_{\Gamma^c} v^\eta(\xi)= \1_{\Gamma^c} v(\xi)
  &\geq\dsE^{\dsQ}\bigg[\1_{\Gamma^c}\Big\{v\big(Z^{\tau, \xi, \pi^{v}}_{\rho\wedge
    \tau_{a}}\big)-\frac{1}{2\varepsilon}\int_{\tau}^{\tau_a\wedge \rho}\Vert \theta_t\Vert^2\df
    t\Big\}\bigg\vert\calG_\tau\bigg]
    \nonumber\\
  &\geq \dsE^{\dsQ}\bigg[\1_{\Gamma^c}\Big\{v^\eta\big(Z^{\tau, \xi, \pi}_{\rho\wedge \tau_{a}}\big)
     -\frac{1}{2\varepsilon}\int_{\tau}^{\tau_a\wedge \rho}\Vert \theta_t\Vert^2\df t\Big\}\bigg\vert\calG_\tau\bigg].\nonumber
\end{align}
 Thus, we deduce that
\begin{align}
  v^\eta(\xi) \geq \dsE^{\dsQ}\bigg[v^\eta\big(Z^{\tau, \xi, \pi}_{\rho\wedge \tau_{a}}\big)
  -\frac{1}{2\varepsilon}\int_{\tau}^{\tau_a\wedge \rho}\Vert \theta_t\Vert^2\df t\bigg|\calG_\tau\bigg].
  \label{upto.min2}
\end{align}
To replace $\rho\wedge\tau_a$ with $\rho$, consider $\Lambda \coloneqq\{\rho > \tau_a\}\in \calG_{\tau_a \wedge \rho}$.
Recall that $v = v^\eta$ at
$\partial B_a(\zf_0)$ and on $ \rrbracket \tau_a, \rho\rrbracket \cap (\Lambda\times\dsR_+) $, we have
$\pi = \pi^{v, a}$. It then follows that
\begin{align}
  \1_\Lambda v^\eta(\xi_a) = \1_\Lambda v(\xi_a)
  &\geq \dsE^{\dsQ} \bigg[\1_\Lambda \Big\{v(Z^{\tau, \xi, \pi^{v, a}}_\rho)
    -\frac{1}{2\varepsilon}\int_{\tau_a}^{\rho}\Vert \theta_t\Vert^2\df t\Big\}\bigg|\calG_{\tau_a}\bigg]\nonumber\\
  &\geq \dsE^\dsQ\bigg[\1_\Lambda \big\{v^\eta(Z^{\tau, \xi, \pi}_\rho)
     -\frac{1}{2\varepsilon}\int_{\tau_a}^{\rho}\Vert \theta_t\Vert^2\df t\Big\}\bigg|\calG_{\tau_a}\bigg]. \label{on.lambda.s}
\end{align}
By \cref{on.lambda.s}, one can derive that
\begin{align}
\dsE^{\dsQ}  \bigg[\1_\Lambda \Big\{v^\eta(\xi_a)-\frac{1}{2\varepsilon}\int_{\tau}^{\tau_a}\Vert \theta_t\Vert^2\df t
     \Big\}\bigg|\calG_{\tau}\bigg]
  &= 
    \dsE^{\dsQ}  \bigg[\dsE^{\dsQ}\bigg[\1_\Lambda
    \Big\{v^\eta(\xi_a)-\frac{1}{2\varepsilon}\int_{\tau}^{\tau_a}\Vert \theta_t\Vert^2\df t
    \Big\}\bigg|\calG_{\tau_a}\bigg]\bigg|\calG_{\tau}\bigg] \nonumber\\
  &\geq\dsE^{\dsQ}  \bigg[\1_\Lambda \Big\{v^\eta(Z_\rho^{\tau, \xi, \pi})-\frac{1}{2\varepsilon}\int_{\tau}^{\rho}\Vert \theta_t\Vert^2\df t
    \Big\}\bigg|\calG_{\tau}\bigg]. \label{on.lambda.s2}
\end{align}
Therefore, thanks to \cref{on.lambda.s}  and  \cref{on.lambda.s2}, the inequality holds that
\begin{align}
  \1_{\Lambda}v^\eta(\xi) \geq \1_{\Lambda}\dsE^{\dsQ}\bigg[v^\eta\big(Z^{\tau, \xi, \pi}_{\rho}\big)
  -\frac{1}{2\varepsilon}\int_{\tau}^{\rho}\Vert \theta_t\Vert^2\df t\bigg|\calG_\tau\bigg].
  \label{upto.min3}
\end{align}
We can obtain the inequality on $\Lambda^c$ in the similar fashion as in the proof of
\Cref{thm.v-}. Hence, it can be shown that $v^\eta\in \calV^+$, which contradicts \cref{calFB.sup} and our claim holds.
\end{proof}
\end{theorem}

\subsection{Proof of Comparison Principle}          
\label{sec:app.cp}
Comparison principle with either Neumann or oblique-type boundary conditions was
already studied; see, for example,
\cite{barles1999nonlinear,barles2014neumann}.   
However, because we have both Dirichlet and oblique-type 
boundary conditions in our problem, some tailor made arguments need to be
developed here. 

We plan to apply a typical doubling argument, nevertheless, the additional
difficulty by considering oblique-type conditions is that we need to construct a
test function with care. We will choose a test function in a way that
$\calB\not=0$ in a viscosity sense. Then by the definition of viscosity
solution, the test function should satisfy $\calF=0$ and this in turn will
provide a contradiction. 
In what follows, we denote
  $\qf^1 \coloneqq [q^1, ~-1-q^1, 0]^\top$ and $\qf^2 \coloneqq [q^2,~0,  ~-1-q^2]^\top$.

To explain the idea to choose a test function, let $\zf, ~\zf' \in \dsR^3$. As
always, to push the variables into a diagonal entry, we need
$\Vert\zf -\zf'\Vert^2/\alpha$ for some $\alpha >0$, in the test function. Moreover, since the
domain $\calO$ is not bounded, for the test function to have a maximum in a
compact set, one may want to put $\beta(\Vert\zf\Vert^2+\Vert\zf'\Vert^2)/2$ for some
$\beta>0$. If we stop here, the test function may or may not satisfy $\calB^i$,
$i \in \{1, 2\}$. To be more precise, for $\zf\in \partial\calO^0$ or
$\zf' \in \partial\calO^0$, we cannot guarantee that
\begin{align}
  \nabla \Big[\frac{1}{\alpha}\Vert\zf -\zf'\Vert^2 +\frac{\beta}{2}(\Vert\zf\Vert^2+\Vert\zf'\Vert^2)\Big]\cdot \qf^i > 0,
  ~~~i \in \{1, 2\}. \label{patho}
\end{align}
To eliminate  the possibility to satisfy $\calB^i$, i.e., to focus on $\calF$, we
seek to remedy the test function to meet \cref{patho}. To this end, pick any
$\nu^i>0$, $i \in \{1, 2\}$, and choose $\zf_\nu \coloneqq (R, \nu^1, \nu^2)$. Then for
any $\zf = (x, 0, y^2) \in \partial\calO^0_1$, we have
$(\zf-\zf_\nu) \cdot \qf^1 = (x-R)q^1 + \nu^1(1+q^1)>0$. Likewise, we also have
$(\zf-\zf_\nu) \cdot \qf^2 >0$ for any $\zf \in \partial\calO^0_2$. 
Therefore, instead of
$\beta(\Vert\zf\Vert^2+\Vert\zf'\Vert^2)/2$, we put
\begin{align}
\chi_\beta(\zf, \zf')\coloneqq 
   \frac{\beta}{2}\Vert \zf-\zf_\nu\Vert^2 +\frac{\beta}{2}\Vert \zf'-\zf_\nu\Vert^2.\nonumber
\end{align}
However, the effect of \cref{patho} is offset by the derivative of
$\Vert\zf -\zf'\Vert^2/\alpha$. Thus, to remove the derivative, we add
additional terms and define
\begin{align}
  \zeta_\alpha(\zf, \zf') \coloneqq
  &\frac{\Vert\zf-\zf'\Vert^2}{2\alpha} + \SUM_{i\in \{1, 2\}}\Big\{C^i_\alpha(\zf, \zf')[d^i(\zf) - d^i(\zf')]
    +\frac{\Vert\qf^i \Vert^2}{2 \alpha (\nf^i \cdot \qf^i)^2}[d^i(\zf) - d^i(\zf')]^2\Big\}, \nonumber\\
  &+\frac{q^1q^2}{2\alpha(1+q^1)(1+q^2)}\big[\SUM_{i \in \{1, 2\}}\{d^i(\zf)-d^i(\zf')\}\big]^2, \nonumber\\
  C^i_\alpha(\zf, \zf')\coloneqq&
   (\zf - \zf') \cdot \qf^i / (\alpha\nf^i \cdot \qf^i),\nonumber\\
  d^i(\zf)\coloneqq &\text{dist}(\zf, \partial\calO^0_i),\nonumber\\
  \nf^1\coloneqq&[0, -1, 0]^\top, \quad\nf^2\coloneqq[0, 0, -1]^\top.\nonumber
\end{align}
Note that $\nabla d^i = -\nf^i$, $i\in \{1, 2\}$, $\nf^1 \cdot \qf^2= \nf^2 \cdot
\qf^1=0$, and
\begin{align}
  \qf^i \cdot \qf^j
  &=
  \begin{cases}
    q^1q^2,~~& i\not=j, \\
    \Vert q^i \Vert^2, &i=j,
  \end{cases}\label{formulaqq}
  \\
  \nf^i \cdot \qf^j
  &=
  \begin{cases}
    0,~~& i\not=j, \\
    1+q^i, &i=j,
  \end{cases}\label{formulanq}        
  \\
  C^i_\alpha(\zf, \zf')\nf^i\cdot \qf^j
  &=
  \begin{cases}
    0,~~& i\not=j, \\
    \alpha^{-1}(\zf - \zf') \cdot \qf^j, &i=j.
  \end{cases}\label{formulac}
\end{align}

Then we define
$\Psi_{\alpha, \beta} \colon \dsR^3 \times\dsR^3 \to \dsR$ as
\begin{align}
  \Psi_{\alpha, \beta}(\zf, \zf')\coloneqq
  &u(\zf) - v(\zf') - \psi _{\alpha, \beta}(\zf, \zf'), \label{def.test}\\
    \psi_{\alpha, \beta}(\zf, \zf')\coloneqq
  &  \zeta_\alpha(\zf, \zf') +\chi_\beta(\zf, \zf'). 
\end{align}
Now, we check some properties of $\psi$ by straightforward calculations. First, we can derive that
\begin{align}
  \nabla_{\zf}\psi_{\alpha, \beta}(\zf, \zf')
  &=\alpha^{-1}(\zf - \zf')+\SUM_{i\in \{1, 2\}}\Big\{
    -C^i_\alpha(\zf, \zf')\nf^i+\qf^i(\alpha \nf^i \cdot \qf^i)^{-1}[d^i(\zf)-d^i(\zf')]
  \nonumber\\
  &-\frac{\Vert\qf^i \Vert^2}{\alpha (\nf^i \cdot \qf^i)^2}[d^i(\zf)-d^i(\zf')]\nf^i\Big\}
    + \beta (\zf-\zf_\nu)\nonumber\\
  &-\frac{q^1q^2}{\alpha(1+q^1)(1+q^2)}
    \big[\SUM_{i \in \{1, 2\}}\{d^i(\zf)-d^i(\zf')\}\big][\nf^1+\nf^2],
    \label{test.first.x}\\    
  \nabla_{\zf'}\psi_{\alpha, \beta}(\zf, \zf')
  &=\alpha^{-1}(\zf' - \zf)+\SUM_{i\in \{1, 2\}}\Big\{
    C^i_\alpha(\zf, \zf')\nf^i-\qf^i(\alpha \nf^i \cdot \qf^i)^{-1}[d^i(\zf)-d^i(\zf')]
    \nonumber\\
  &+\frac{\Vert\qf^i \Vert^2}{\alpha (\nf^i \cdot \qf^i)^2}[d^i(\zf)-d^i(\zf')]\nf^i\Big\}+ \beta
    (\zf'-\zf_\nu) \nonumber\\
  &+\frac{q^1q^2}{\alpha(1+q^1)(1+q^2)}
    \big[\SUM_{i \in \{1, 2\}}\{d^i(\zf)-d^i(\zf')\}\big][\nf^1+\nf^2].
    \label{test.first.y}
\end{align}
Moreover, we can observe that
\begin{align}
  \nabla_\zf\zeta_\alpha(\zf, \zf') =& -\nabla_{\zf'}\zeta_\alpha(\zf, \zf'), \nonumber\\
  \nabla_\zf\psi_{\alpha, \beta}(\zf, \zf') =& -\nabla_{\zf'}\psi_{\alpha, \beta}(\zf, \zf') + \beta(\zf-\zf_\nu)+\beta(\zf'-\zf_\nu).\nonumber
\end{align} 
Hence, recalling \cref{formulaqq}-\cref{formulac}, for any $\zf\in \calO $, $i
\not=j$, $\zf_j \in \partial\calO^0_j$, we have 
\begin{align}
  \nabla_{\zf}\psi_{\alpha, \beta}(\zf_j, \zf)\cdot \qf^j
  &= \beta(\zf_j-\zf_\nu)\cdot \qf^j +\frac{q^1q^2}{\alpha(1+q^i)}d^j(\zf)  >0, \label{app.test.boundary.x}\\
  \nabla_{\zf'}(-\psi_{\alpha, \beta})(\zf_j, \zf)\cdot \qf^j
  &= -\beta(\zf_j-\zf_\nu)\cdot \qf^j -\frac{q^1q^2}{\alpha(1+q^i)}d^j(\zf)<0.
 \label{app.test.boundary.y}
\end{align}
(\ref{app.test.boundary.x})-(\ref{app.test.boundary.y}) will be used later in the 
proof of \Cref{thm.comp}. In addition, from (\ref{test.first.x}) - (\ref{test.first.y})
, the second order derivative of $\psi$ is obtained. Let
\begin{align}
  A \coloneqq \If_3+\SUM_{i \in \{1,2,\}}\Big\{
  \frac{\Vert\qf^i \Vert^2\nf^i (\nf^i)^\top}{(\nf^i \cdot \qf^i)^2}-\frac{\nf^i(\qf^i)^\top
  + \qf^i (\nf^i)^\top}{(\nf^i \cdot \qf^i)}\Big\}
  +\frac{q^1q^2}{(1+q^1)(1+q^2)}[\nf^1+\nf^2][\nf^1+\nf^2]^\top,\nonumber
\end{align}
where $\If_3$ is the $3\times3$-identity matrix. If $q^i$, $i \in \{1,2\}$, are not too
big, we clearly have $A \succeq 0$. Then we
can write
\begin{align}
  \nabla^2\psi_{\alpha, \beta}(\zf, \zf') =\frac{1}{\alpha}
  \begin{bmatrix}
    A &-A\\
    -A & A
  \end{bmatrix} 
         +\beta\begin{bmatrix}
           \If_3&0 \\
           0 & \If_3
         \end{bmatrix}.\nonumber
\end{align}
We are ready to prove the comparison principle.
\begin{proof}[Proof of \Cref{thm.comp}]
We argue by contradiction. To this end, we suppose that for some
$\zf_e \in \clco$, $u(\zf_e) - v(\zf_e) = \delta >0$. Let us choose 
$\beta$ small enough such that
 $ \delta > \chi_\beta(\zf_e, \zf_e)$, 
and choose $\{\alpha_n\}_{n\in \dsN}$ such that $\alpha_n \downarrow0$. Denote
 $\Psi_n\coloneqq \Psi_{\alpha_n, \beta}$. 
As $u$ and $v$ are bounded, $\chi_\beta$ dominates $u-v$ outside a compact
set. Therefore, for each $n \in \dsN$, $\Psi_{n}$ has its maximum on
$\clco\times\clco$ in a compact set and we denote the maximal point by
$(\zf_n, \zf_n')$, i.e., 
\begin{align} 
  \Psi_n(\zf_n, \zf_n') =
  &\sup_{(\zf, \zf')\in\clco\times\clco}\Psi_n(\zf, \zf').\nonumber
 \end{align}
The maximal point $(\zf_n, \zf'_n)$ actually depends on $\beta$ but we drop it for
simplicity. As $\{(\zf_n, \zf'_n)\}_{n\geq1}$ lie in a compact set, we choose a
convergent subsequence, still denoted by $(\zf_n, \zf_n')$, such that
\begin{align}
  (\zf_n, \zf_n')\to (\zfb, \zfb') = (\overline{x}, \overline{y}, \overline{x}',\overline{ y}').\nonumber
\end{align}
As $u \leq v$ on $\partial\calO_R \cup \partial\calO_{c/r}$ by the definition of viscosity sub/super
solution, $(\zfb, \zfb')$ must be in $\calO\times
\calO$. The previous assumption yields that
\begin{align}
  \Psi_n(\zf_n, \zf_n') \geq \sup_{\zf\in \clco}
  [u(\zf) - v(\zf)  -\chi_\beta(\zf, \zf)] \geq \delta - \chi_\beta(\zf_e, \zf_e) >0.\nonumber
\end{align}
Therefore, it follows that
\begin{align}
  \zeta_{\alpha_n}(\zf_n, \zf'_n) \leq u(\zf_n) - v(\zf'_n)  -\chi_\beta(\zf_n, \zf'_n)
  -\sup_{\zf\in \clco} [u(\zf) - v(\zf)  -\chi_\beta(\zf, \zf)].\nonumber
\end{align}
In view that the right hand side is bounded above but $\alpha_n\to 0$ as $n\to \infty$,
 $(\overline{x}, \overline{y}) = (\overline{x}', \overline{y}')$. 
Moreover, the fact that $u-v$ is u.s.c implies that
\begin{align}
  0\leq  \limsup_{n\to\infty} \zeta_{\alpha_n}(\zf_n, \zf'_n) \leq& u(\zfb) - v(\zfb')
  -\chi_\beta(\zfb, \zfb')  
  - \sup_{\zf\in \clco} [u(\zf) - v(\zf)  -\chi_\beta(\zf, \zf)] \leq0.\nonumber
\end{align}
Hence, $\lim_{n\to \infty}\zeta_{\alpha_n}(\zf_n, \zf'_n)=0$.

By Crandall-Ishii's lemma, for large $n\in \dsN$, there exist $\Af_n, \Bf_n \in  
\calS^3$ such that
\begin{align}
(\nabla_{\zf}\psi_{\alpha_n, \beta}(\zf_n,\zf'_n),  \Af_n) \in \overline{\calJ}^{2, +}_{\calO} u(\zf_n), \quad\quad
(-\nabla_{\zf'}\psi_{\alpha_n, \beta}(\zf_n,\zf'_n), \Bf_n) \in \overline{\calJ}^{2, -}_{\calO} v(\zf'_n)\nonumber
\end{align}
and that
  \begin{align}\label{cilemma}
  -\frac{10}{\alpha_n}
  \begin{bmatrix}
    \If_3 & 0\\ 0&\If_3
  \end{bmatrix}
      \prec
  \begin{bmatrix}
    \Af_n & 0 \\0 &-\Bf_n 
  \end{bmatrix}
      \prec\frac{10}{\alpha_n}
  \begin{bmatrix}
        \If_3&-\If_3 \\-\If_3 &\If_3
  \end{bmatrix} + 2\beta \begin{bmatrix}
        \If_3&0 \\0 &\If_3
  \end{bmatrix}. 
  \end{align}
We can calculate that
\begin{align}
  \nabla_{\zf}\psi_{\alpha_n, \beta}(\zf_n,\zf'_n) =& \nabla_{\zf}\zeta_{\alpha_n}(\zf_n,\zf'_n) + \beta(\zf_n-\zf_\nu)
  \nonumber\\
  -\nabla_{\zf'}\psi_{\alpha_n, \beta}(\zf_n,\zf'_n) =& \nabla_{\zf}\zeta_{\alpha_n}(\zf_n,\zf'_n) - \beta(\zf'_n-\zf_\nu).\nonumber
\end{align}
Let $F$ be the function such that
$\calF[\varphi](\zf) = F(\zf, \varphi(\zf), \nabla \varphi(\zf), \nabla^2 \varphi(\zf))$.
Then we have
\begin{align}
 \lambda^D(u(\zf_n) -v(\zf'_n))
  =& F(\zf_n, u(\zf_n), \nabla_{\zf}\psi_{\alpha_n, \beta}(\zf_n,\zf'_n), \Af_n) 
     -F(\zf_n, v(\zf'_n),  \nabla_{\zf}\psi_{\alpha_n, \beta}(\zf_n,\zf'_n), \Af_n) \nonumber\\
  \leq&F(\zf'_n, v(\zf_n), \nabla_{\zf}\psi_{\alpha_n, \beta}(\zf_n,\zf'_n), \Bf_n) 
     -F(\zf_n, v(\zf'_n),  \nabla_{\zf}\psi_{\alpha_n, \beta}(\zf_n,\zf'_n), \Af_n) \nonumber\\
  \leq&F\big(\zf'_n, v(\zf_n), \nabla_{\zf}\zeta_{\alpha_n, \beta}(\zf_n,\zf'_n),
     \Bf_n+2\beta\If_3\big) \nonumber\\
   &\quad\quad -F\big(\zf_n, v(\zf'_n),  \nabla_{\zf}\zeta_{\alpha_n, \beta}(\zf_n,\zf'_n),
     \Af_n-2\beta\If_3\big) + c(\beta), \label{comp.cal1}
\end{align}
where $c(\beta)$ is the modulus of continuity of $F$. The last inequality of 
\cref{comp.cal1} is obtained by the compactness of $\calK$. By \cref{cilemma},
we moreover, have
 $\Af_n - 2\beta\If_3 \prec \Bf_n + 2\beta\If_3$.
 Therefore, we obtain
\begin{align}
  F\big(\zf'_n, v(\zf_n), &\nabla_{\zf}\zeta_{\alpha_n, \beta}(\zf_n,\zf'_n),
     \Bf_n-2\beta \If_3\big) \nonumber\\
   &\quad\quad \leq F\big(\zf_n, v(\zf'_n),  \nabla_{\zf}\zeta_{\alpha_n, \beta}(\zf_n,\zf'_n),
     \Af_n+2\beta \If_3\big). \label{comp.cal2}
\end{align}
By \cref{comp.cal1} and \cref{comp.cal2},  taking $n \uparrow \infty$ leads to
 $ \lambda^D \delta \leq c(\beta)$. 
Again taking $\beta \downarrow 0$, we have the desired contradiction, which completes the proof. 
\end{proof}
\subsection{Proof of \Cref{thm.main}}

Finally, we are ready to prove our main result of \Cref{thm.main}.
\begin{proof}[Proof of \Cref{thm.main}] 
\Cref{thm.v-}, \Cref{thm.v+}, together with
\Cref{thm.comp} imply that
  $v^+ \leq v^-$. Therefore, it suffices to show
 $v^- \leq V \leq v^+$. 
To show the first inequality, let us consider an arbitrary $\phi \in \calV^-$. It is obvious that
$\phi \leq V$ on $\partial\calO_R\cup\partial\calO_{c/r}$. Let $(x, y) \in \calO$ and take the
\textit{random initial condition} as $\tau=0$ and $\xi=(x, y)$. We fix some $\pi \in \scrR$ and the hitting time defined by
\begin{align}
  \tau^{\tau, \xi, \pi}_{c/r} \coloneqq& \inf\{ t \geq0 \colon X^{\tau, \xi, \pi}_t \geq
                               c/r\}.\nonumber 
\end{align}
As there exists $\theta$-generated measure $\dsQ$ for $\phi$ to satisfy (SB2), it
follows that 
\begin{align}
  \phi(x, y) &\leq
  \dsE^\dsQ\bigg[\phi(Z^{\tau, \xi, \pi}_{\tau^{\tau, \xi, \pi}_E})
  -\frac{1}{2a}\int_{\tau}^{\tau^{\tau, \xi, \pi}_E}e^{-\lambda^D s}\Vert\theta_s \Vert^2\df s~\bigg|~ \calG_\tau\bigg]\nonumber\\
   &\leq \dsE^\dsQ\bigg[\1_{\tau^{\tau, \xi, \pi}_E = \tau^{x, y, \pi}_R}
  -\frac{1}{2a}\int_{\tau}^{\tau^{\tau, \xi, \pi}_E}e^{-\lambda^D s}\Vert\theta_s \Vert^2\df s~\bigg|~ \calG_\tau\bigg]. \label{just1}
\end{align}
Moreover, we have
\begin{align}
  \dsE^\dsQ[\1_{\tau^{\tau, \xi, \pi}_E = \tau^{x, y, \pi}_R}]
  = \dsQ[\tau^{\tau, \xi, \pi}_R < \tau_D \wedge \tau^{x, y, \pi}_{c/r}] \leq \dsQ[\tau^{x, y, \pi}_R < \tau_D].\label{just2}
\end{align}
By combining \cref{just1} and \cref{just2}, we have   
$\phi(x, y) \leq  V(x, y)$, together with \cref{sup.sto.sub} yield $ v^- \leq V$.  In a similar fashion, we can show $V \leq v^+$ as well.
Because $v^-$ is a viscosity super-solution, by \Cref{thm.comp}, we have
$v^+ \leq v^-$. It follows that $v^- \leq V \leq v^+\leq v^-$, which readily implies our desired equality $v^- = V =v^+$ and hence the value function is the unique viscosity
  solution of the HJB equation \cref{dpe.crude}.
\end{proof}

\section*{Acknowledgments}
The first and third authors acknowledge the support from the Singapore MOE AcRF
grants R-146-000-271-112 and R-146-000-255-114 as well as the French Ministry of
Foreign Affairs and the Merlion programme.  The second author is partially
supported by the Hong Kong Early Career Scheme under grant no. 25302116 and the
Hong Kong Polytechnic University central research grant under no.15304317. In
addition, the first author received the financial support from the Singapore MOE
AcRF grant R-146-000-243-114 and the third author received the financial support
from the NSFC Grant 11871364.

\bibliographystyle{abbrv}       
\bibliography{hwm}   

\end{document}